\documentclass[12pt,a4paper]{article}
\usepackage{amsmath,amssymb, amsfonts, amsthm}
\usepackage{graphicx}

\begin{document}

\title{{\Large \textbf{A model of adaptive decision making from
representation of information environment by quantum fields}}}
\author{ F. Bagarello \\
DEIM, Facolt\`{a} di Ingegneria,\\
Universit\`{a} di Palermo, I - 90128 Palermo, and\\
INFN, Sezione di Napoli, Italy.\\
E-mail: fabio.bagarello@unipa.it\\
home page: www1.unipa.it/fabio.bagarello \\
E. Haven\\
School of Business and IQSCS\\
University of Leicester, LE1 7RH, UK\\
A. Khrennikov\\
International Center for Mathematical Modeling\\
in Physics and Cognitive Scienc\\
Linnaeus University, S-35195, V\"{a}xj\"{o}, Sweden, and\\
National Research University of  Information Technologies,\\
Mechanics and Optics (ITMO), St. Petersburg 197101, Russia}

\maketitle

\begin{abstract}
We present the mathematical model of decision making (DM) of agents acting
in a complex and uncertain environment (combining huge variety of
economical, financial, behavioral, and geo-political factors). To describe
interaction of agents with it, we apply the formalism of quantum field
theory (QTF). Quantum fields are of the purely informational nature. The
QFT-model can be treated as a far relative of the expected utility theory,
where the role of utility is played by adaptivity to an environment (bath).
However, this sort of utility-adaptivity cannot be represented simply as a
numerical function. The operator representation in Hilbert space is used and
adaptivity is described as in quantum dynamics. We are especially interested
in stabilization of solutions for sufficiently large time. The outputs of
this stabilization process, probabilities for possible choices, are treated
in the framework of classical DM. To connect classical and quantum DM, we
appeal to Quantum Bayesianism (QBism). We demonstrate the quantum-like
interference effect in DM which is exhibited as a violation of the formula
of total probability and hence the classical Bayesian inference scheme.
\end{abstract}

\section{Introduction}

This paper is devoted to applications of the mathematical formalism of
quantum theory to the modeling of decision making (henceforth denoted as
`DM') in a context where there exists deep uncertainty\footnote{%
We will discuss below what we mean with `deep uncertainty'.} about both the
possible actions of other decision makers and the surrounding complex
\textit{information environment. }This information environment can include
belief-states of decision makers, memory recollections, and external
economic (or also financial, social and geopolitical) contexts\footnote{%
The notion `state of the world'\ used in economics (and decision making) and
financial asset pricing can be considered as the counterpart of these
physical notions (see Machina \cite{Machina2003}).}. In economics, the
information environment is typically formalized via the introduction of the
notions of i) public (non-scarce) and private (scarce) information (flows);
and the notions of ii) produced information (a data-set which needs buying
for instance) and emergent information (time which reveals information for
instance). The economics and finance literature has published research on
how one can measure the economic value of information. The notion of
`private information price of risk' for instance measures how private
information triggers changes in the price of risk within an asset pricing
context (see Detemple and Rindisbacher \cite{Detemple2013}, \cite%
{Detemple2016}). The idea of `degrees of information' is sometimes also
considered such as `first-order' information (which ties the information to
an event) and `higher-order' information (which refers to the information
itself (for instance what is the source of the information)). The notion of
`common knowledge' which we briefly mention below is an example of such
higher-order information. For a detailed discussion of many of the basic
concepts (see \cite{Bikh2013}, p. 171- and p.224-).

The idea of `information environment' is also used in physics and very often
the terms `bath'\ or `reservoir'\ are used. In a social science adapted
context (starting from physics) we can then speak of `mental bath' or
`mental reservoir'.

In the model presented here in this paper, agents are not aiming to maximize
their utility, but they instead adapt their behavior to information gained
from other agents and the environment. To describe this process of
adaptivity, we adopt the mathematical apparatus of Quantum Field Theory
(QFT). In particular, the information environment is represented
mathematically by quantum fields. We call this approach \textit{QFT-inspired}%
. It has to be sharply distinguished from a variety of applications where
genuine quantum physics is applied to cognition and DM.

The adaptive approach which we will introduce, is more realistic compared to
the utility maximizing stance used in basic economics. However, the ensuing
formalism is much more demanding. As is well known, from a basic economics
perspective, the usual assumption made is that in the absence of any
arbitrage\footnote{%
An arbitrage opportunity can be formally defined (see Bj\"{o}rk \cite%
{Bjork1998}). Essentially, an arbitrage opportunity arises when no
investment needs making to obtain a positive cash flow. This is akin to a
situation where a positive return on an investment exists without the
investment itself having any risk of its own. From a risk-reward
perspective, there is positive reward for no risk.} opportunities there
should exist a so called budget feasible plan which can not be more
preferred by any other budget feasible plan (see Ma \cite{Ma2011}).

From the brief discussion so far, it is intuitive, that our DM model will
differ crucially from the classical expected utility theory introduced by
von Neumann and Morgenstern \cite{vonNeumann1944}. Nevertheless, the
QFT-inspired adaptive model can be treated as a far relative of utility
based models of DM, including expected utility. The role of utility is
played by the degree of adaptivity to the surrounding (informational)
environment, i.e. the mental bath. However, it is impossible to encode
`environment-adaptivity' by a numerical utility-function. As we will show,
it is encoded by the dynamics of quantum decision operators.

Within the mainstream literature, in cognitive psychology and behavioral
economics, several models have been developed to measure certain effects
such as `order' effects and `conjunction' and `disjunction' effects. Those
are effects which have been repeatedly observed in experimental lab settings
and some of those effects have relevance to some of the important paradoxes
which have affected the integrity of the axiomatic structure of key expected
utility models (such as the von Neumann-Morgenstern expected utility model
(Allais' paradox, see \cite{Allais1953}) and the Savage expected utility
model (Ellsberg paradox, see \cite{Ellsberg1961}). See also Machina \cite%
{Machina2009}). A very rich literature has emerged and we must mention the
work of Tversky and Kahneman \cite{Tversky1983} and Tversky and Shafir \cite%
{Tversky1992}. However, we need to note that those effects are now also
modelled with a so called `quantum-like' approach. In essence, this approach
uses formalisms from quantum mechanics to augment the modelling of
non-physics problems, such as cognition. For representative work in this
area see \cite{Aerts2012}-\cite{Trueblood2012} and references therein.

Finally, the quantum-like approach has recently been successfully applied to
the theory of common knowledge and the modeling of violations of Aumann's
theorem about the impossibility to agree on disagreeing (see Khrennikov \cite%
{Khrennikov2015} and \cite{Khrennikov2014}).

As was already pointed out, we are interested in modeling the DM-process as
being adaptive to the information environment. An agent $\mathcal{G}$ is not
driven to maximize his utility, but he rather tries to adapt his behavior to
information which he is gaining (during the DM-process) from his environment
$\mathcal{R}.$

There are two well known main approaches to the modeling of quantum adaptive
dynamics. One approach is based on the theory of \textit{open quantum
systems.} Here the state of an agent is extracted from the general state of
the compound system `agent+environment' and the dynamics are reduced (by
using the trace operation) to the state space of this agent (the environment
is encoded in the operator-coefficients of the quantum master equation)%
\footnote{%
This approach was used for a wide class of problems in DM, psychology,
politics and biology (see \cite{Asano2011}-\cite{Asano2015}; \cite{Haven2013}%
; \cite{Khrennikova2016}).}. Another approach is based on the study of the
general dynamics of the state of the compound system, `agent+environment',
and then averages and probabilities corresponding to the agents' decisions
are extracted from the complete dynamical state (see \cite{Bratteli1987}).
In the latter approach an environment is represented as a quantum field. In
this paper we proceed with this QFT-inspired approach by using the
mathematical methods developed in recent papers by Bagarello \cite%
{Bagarello2015a}-\cite{Bagarello2016} and in the monograph \cite%
{Bagarello2012}.

The main advantage of the open quantum system approach is the reduction of
the state-dimension to the dimension of the agent's state. The main problem
is that an environment appears in a very formal encoding and it is not easy
to extract its features from the coefficients of the quantum master
equation. Thus, \textit{the model is phenomenological}\footnote{%
Another complication is that it is impossible to represent the quantum
adaptive dynamics as dynamics of pure states. The impact of the environment
can transfer a pure initial state into a mixed quantum state (represented by
a density operator).}.

The QFT-inspired approach is more difficult analytically, since the
scenarios of evolution are presented by taking into account all degrees of
freedom of the environment\footnote{%
Such an environment contains both the complex internal mental
representations of agents as well as, e.g. the market situation or the
current state of the political arena.}. It can be difficult to construct
analytical solutions of the dynamical equations for the state evolution,
even if in many relevant cases this can be done (see sections \ref{TMOD} and %
\ref{sectnumres}). Sometimes, however, numerical methods are really what is
needed. This is the case, for instance, of non-linear models, which will not
play any role in this paper. Amongst the advantages of the QFT-inspired
approach, we can point to the possibility of describing dynamics in terms of
solely pure states.

At the beginning of this paper, we mentioned the idea of `deep uncertainty'.
From the quantum information point of view (see Chiribella et al. \cite%
{Chiribella2012}) a pure state represents the maximally available
information about the context (in our case the context of DM). But, the
availability of maximal information does not mean the resolution of
uncertainty. We say that the uncertainty can be very deep. The pure state
representation means that maximal information (about the context) is encoded
in it. The most tricky point of quantum information theory (and, in fact,
quantum theory in general) is that there exist states of compound systems,
e.g. `agent+environment', which are pure. This means they encode maximal
information, but the states of subsystems, e.g., the state of the agent, are
not pure (i.e., they do not encode maximally available information about the
situation). Such states of compound systems are called \textit{entangled}.
Entanglement is nowadays considered as the main distinguishing feature of
quantum theory.

Let us now come back to the objective of this paper. We will study a very
general DM-model represented in the form of a game played by two players $%
\mathcal{G}_{1}$ and $\mathcal{G}_{2}$ interacting with two environments, $%
\mathcal{R}_{1}$ and $\mathcal{R}_{2}.$ Such players can be agents of a
market, e.g. corporations, traders of the financial market, political
parties. A decision is taken only after the interactions $\mathcal{G}%
_{j}\leftrightarrow \mathcal{R}_{j}$ and $\mathcal{G}_{1}\leftrightarrow
\mathcal{G}_{2}$ are considered. The reflections of the agents generated by
the interactions modify the agents' mental states. Thus, as was already
emphasized, the key-notion of such a decision making model is not \textit{%
utility}, but rather the \textit{interactions between agents and their
environments and the adaptivity to their impact}\footnote{%
It is important to stress that in quantum theory `interactions'\ cannot be
imagined as actions of forces, as we known them in classical physics. For
our applications, the most useful is the information interpretation of
quantum theory (See Chiribella et al. \cite{Chiribella2012}; Brukner and
Zeilinger \cite{Brukner1999}). See Plotnitsky \cite{Plotnitsky2014} for
applications of the information interpretation of quantum theory to DM.}.

The QFT-inspired model for this sort of games involving decision making
under uncertainty (including the uncertainty generated by a complex
surrounding environment (be it mental, economic, social,...), was considered
in papers by \cite{Bagarello2015a}, \cite{Bagarello2015b}. In these papers
one of the main characteristics of the analysis was that, at the beginning
of the process of decision making $t=0$, each agent was in a \emph{sharp
state}. With this we mean that each agent knew, at $t=0$, exactly which was
her own choice. Thus, initially, an agent was able to resolve her internal
mental uncertainty. Then, in the process of decision making, her certainty
was destroyed as the result of reflections generated by interactions with
another agent and the environment.

Of course, this initial certainty of the agent's state is a strongly
simplifying assumption which does not match reality (e.g. when we consider
traders in the financial market).

In this paper we present a more realistic model by considering the
possibility that, initially, an agent is also being in a state of
uncertainty which is mathematically represented as a superposition of
quantum(-like) pure states corresponding to the concrete choices. Such a
superposition encodes a very deep uncertainty which cannot be modeled in the
framework of classical probability (CP) (see \cite{Busemeyer2012}; \cite%
{Khrennikov2010}). The superposition of decisions induces a kind of
interference between possible decisions of agents (see section \ref%
{sectnumres}). In probabilistic terms, this interference is visible as a
violation of \textit{the classical formula of total probability:} it is
perturbed by an additional term. We will see that, in comparison with \cite%
{Bagarello2015a}, \cite{Bagarello2015b} and \cite{Bagarello2016}, the
dynamical behavior of \textit{the decision functions} (DFs, see below)
changes drastically. In particular, a sort of overall noise appears because
of the presence of some interference effects in the system.

\section{Why quantum(-like)?}

\label{WQL}

The original general idea of Bagarello \cite{Bagarello2012} was to formally
use creation and annihilation operators, the basic building blocks of QFT,
in macroscopic systems so one can possibly consider dynamical systems. These
operators can be used to represent very complex dynamical processes composed
of elementary acts of creation and annihilation of new states of systems.
These states can be of any nature: physical, mental, biological. This is the
minimal interpretation of the QFT-inspired approach. From this formal
viewpoint, the application of the mathematical formalism of quantum theory
is just a matter of convenience, and in fact it proved to be useful in
several applications in different contexts \cite{Bagarello2012}-\cite%
{Bagarello2016}.

In our concrete model the states under consideration are possible strategies
of players $\mathcal{G}_{1}$ and $\mathcal{G}_{2}.$ \textit{The creation and
annihilation operators are explored to model the process of how players
reflect upon their strategies. }These reflections are composed of elementary
acts of transition of the player's mental state from one strategy to another
(\textquotedblleft annihilation\textquotedblright\ of the previously chosen
strategy and \textquotedblleft creation\textquotedblright\ of the opposite
strategy). We remark that this model, as well as the quantum physical model
in general, is epistemic, i.e., it does not represent the internal (and
extremely complex) processes in the neural system leading to such
transitions. It provides only the formal operator representation of these
transitions. However, this formal model is rather rich mathematically, see
sections \ref{TMOD}, \ref{sectnumres}. We can make the argument, that
without such an operational reduction of complexity it would be really
impossible to proceed to concrete solutions of dynamical equations. This
operational reduction of complexity is one the most important features
motivating the exploration of the quantum formalism.

However, the application of this formalism has also deeper foundational
consequences. The creation and annihilation operators act on some complex
Hilbert space. Therefore, the states of this space can form superpositions.
In the quantum formalism, the presence of superpositions of states lead to
non-classical probabilistic effects. One of them is the \textit{interference
of probabilities.} In quantum physics this interference effect can be
written as a violation of the formula of total probability. The latter is
one of the basic laws of classical probability.

As was mentioned in the introduction, the consideration of superpositions of
states is very natural for the modeling of mental phenomena. They represent
uncertainty at the level of an individual decision maker. And, for
superpositions, our model exhibits the aforementioned deviation from
classical probability theory, in the form of the violation of the formula of
total probability, see section \ref{INTS2}.

We also remark that the use of superpositions as initial states generates
oscillations, see Figures \ref{fig1}, \ref{fig2} and \ref{fig3}, section \ref%
{sectnumres}, which are absent for dynamics starting with states
representing ``definite strategies'', \cite{Bagarello2015a}.

In short, the use of the quantum formalism for decision making can be
justified as follows:

\begin{itemize}
\item It provides a powerful operational tool for the representation of
reflections of decision makers.

\item It provides the unique possibility to represent uncertainty in the
mental representation of the decision problem of an \textit{individual
decision maker.}\footnote{%
We remark that classical probability theory also can be used to represent
uncertainty. But a probability measure represents uncertainty in an
ensemble, i.e., a statistical uncertainty (and not in individual setting).
It seems that the quantum representation of uncertainty matches the modeling
of uncertainty \textquotedblleft in the head of an
individual\textquotedblright , see section \ref{SPR} for further discussion.}
\end{itemize}

\section{QFT-inspired model of decision making in two players game and its
dynamics}

\label{TMOD}

In this section we will discuss the details of our (QFT-inspired) model,
constructing first the vectors of the players, the Hamiltonian of the
system, and deducing out of it, the differential equations of motion and
their solution, and we will be particularly interested in its asymptotic (in
time $t$) behavior.

In our game we have two players, $\mathcal{G}_{1}$ and $\mathcal{G}_{2}$.
Each player could operate, at $t=0$, two possible choices, 0 and 1. Hence,
we have four different possibilities, to which, following Bagarello \cite%
{Bagarello2015a}, we associate four different and mutually orthogonal
vectors in a four dimensional Hilbert space $\mathcal{H}_{\mathcal{G}}$.
These vectors are $\varphi _{0,0}$, $\varphi _{1,0}$, $\varphi _{0,1}$ and $%
\varphi _{1,1}$. The first vector, $\varphi _{0,0}$, describes the fact
that, at $t=0$, the two players have both chosen 0 ($0_{1}0_{2}$). Of
course, such a choice can change during the time evolution of the system.
Analogously, $\varphi _{0,1}$ describes the fact that, at $t=0$, the first
player has chosen 0, while the second has chosen 1 ($0_{1}1_{2}$). And so
on. $\mathcal{F}_{\varphi }=\{\varphi _{k,l},\,k,l=0,1\}$ is an orthonormal
basis for $\mathcal{H}_{\mathcal{G}}$. The general mental state vector of
the system $\mathcal{S}_{\mathcal{G}}$ (i.e. of the two players), for $t=0$,
is a linear combination
\begin{equation}
\Psi _{0}=\sum_{k,l=0}^{1}\alpha _{k,l}\varphi _{k,l},  \label{20}
\end{equation}%
where it is natural to assume that $\sum_{k,l=0}^{1}|\alpha _{k,l}|^{2}=1$
in order to normalize the total probability. Indeed, for instance, we
interpret $|\alpha _{0,0}|^{2}$ as the probability that $\mathcal{S}_{%
\mathcal{G}}$ is, at $t=0$, in a state $\varphi _{0,0}$, i.e. that both $%
\mathcal{G}_{1}$ and $\mathcal{G}_{2}$ have chosen 0. Analogous
interpretations can be given to the other coefficients.

We construct the states $\varphi_{k,l}$ by using two fermionic operators,
i.e. two operators $b_1$ and $b_2$, satisfying the following \textit{%
canonical anti-commutation rules} (CAR):
\begin{equation}
\{b_k,b_l^\dagger\}=\delta_{k,l}\,1 \!\! 1,\qquad \{b_k,b_l\}=0,  \label{21}
\end{equation}
where $k,l=0,1$. Here $1 \!\! 1$ is the identity operator and $\{x,y\}=xy+yx$
is the anticommutator between $x$ and $y$. Then we take $\varphi_{0,0}$ as
the vacuum of $b_1$ and $b_2$: $b_1\varphi_{0,0}=b_2\varphi_{0,0}=0$, and
build up the other vectors out of it:
\begin{equation*}
\varphi_{1,0}=b_1^\dagger\varphi_{0,0}, \quad
\varphi_{0,1}=b_2^\dagger\varphi_{0,0}, \quad
\varphi_{1,1}=b_1^\dagger\,b_2^\dagger\varphi_{0,0}.
\end{equation*}

\vspace{2mm}

\textbf{Remarks:--} (1) Notice that $\mathcal{F}_\varphi$ could be
equivalently constructed starting from the vector $\varphi_{1,1}\in\mathcal{H%
}$ which is annihilated by $b_1^\dagger$ and $b_2^\dagger$, and then
constructing the other vectors of $\mathcal{F}_\varphi$ by acting on $%
\varphi_{1,1}$ with $b_1$ and $b_2$.

\vspace{2mm}

(2) The reason why we introduce the operators in (\ref{21}) is because they
are used to write down a Hamiltonian-like operator for the system $\mathcal{S%
}$, from which we deduce the dynamics of the \emph{observables} of $\mathcal{%
S}$, i.e. the variables needed to describe $\mathcal{S}$, and $\mathcal{S}_{%
\mathcal{G}}$ in particular, via the Heisenberg rule.

\vspace{2mm}

An explicit representation of these vectors and operators can be found in
many textbooks in quantum mechanics (see \cite{Roman1965}). For instance: $%
\varphi _{k,l}=\varphi _{k}^{(1)}\otimes \varphi _{l}^{(2)}$, where $\varphi
_{0}=\left(
\begin{array}{c}
1 \\
0 \\
\end{array}%
\right) $ and $\varphi _{1}=\left(
\begin{array}{c}
0 \\
1 \\
\end{array}%
\right) $. Then,
\begin{equation*}
\varphi _{1,0}=\varphi _{1}^{(1)}\otimes \varphi _{0}^{(2)}=\left(
\begin{array}{c}
0 \\
1 \\
\end{array}%
\right) \otimes \left(
\begin{array}{c}
1 \\
0 \\
\end{array}%
\right) ,\qquad \varphi _{1,1}=\varphi _{1}^{(1)}\otimes \varphi
_{1}^{(2)}=\left(
\begin{array}{c}
0 \\
1 \\
\end{array}%
\right) \otimes \left(
\begin{array}{c}
0 \\
1 \\
\end{array}%
\right) ,
\end{equation*}%
and so on. The matrix form of the operators $b_{j}$ and $b_{j}^{\dagger }$
are also quite simple. For instance,
\begin{equation*}
b_{1}=\left(
\begin{array}{cc}
0 & 1 \\
0 & 0%
\end{array}%
\right) \otimes \left(
\begin{array}{cc}
1 & 0 \\
0 & 1%
\end{array}%
\right) ,\qquad b_{2}=\left(
\begin{array}{cc}
1 & 0 \\
0 & 1%
\end{array}%
\right) \otimes \left(
\begin{array}{cc}
0 & 1 \\
0 & 0%
\end{array}%
\right) ,
\end{equation*}%
and so on.

Let now $\hat{n}_{j}=b_{j}^{\dagger }b_{j}$ be the number operator of the $j$%
-th player: the CAR above implies that $\hat{n}_{1}\varphi _{k,l}=k\varphi
_{k,l}$ and $\hat{n}_{2}\varphi _{k,l}=l\varphi _{k,l}$, $k,l=0,1$. Then,
because of what we discussed before, the eigenvalues of these operators
correspond to the choice operated by the two players at $t=0$. For instance,
$\varphi _{1,0}$ corresponds to the choice $1_{1}0_{2}$, just because `one'
is the eigenvalue of $\hat{n}_{1}$ and `zero' is the eigenvalue of $\hat{n}%
_{2}$. It is natural, therefore, to call $\hat{n}_{1}$ and $\hat{n}_{2}$ the
\emph{strategy operators} (at $t=0$). Moreover, since $b_{j}$ and $%
b_{j}^{\dagger }$ modify the attitude of $\mathcal{G}_{j}$, they can be
called the \emph{reflection operators}. We repeat that fermionic operators
are in use, since the eigenvalues of $\hat{n}_{j}$ are exactly 0 and 1,
which are the only possible choices of the players of our game.

Our main effort now consists in \emph{giving a dynamics} to the number
operators $\hat{n}_{j}$, following the scheme described in \cite%
{Bagarello2012}. Therefore, what we first need is to introduce a Hamiltonian
$H$ for the system. Then, we will use this Hamiltonian to deduce the
dynamics of the number operators as $\hat{n}_{j}(t):=e^{iHt}\hat{n}%
_{j}e^{-iHt}$, and finally we will compute the mean values of these
operators on some suitable state which describes (see below) the status of
the system at $t=0$. We refer to \cite{Bagarello2012,Bagarello2015a} for the
details of our construction. Here we just recall that $H$ is the Hamiltonian
of an open system, since the two players $\mathcal{G}_{1}$ and $\mathcal{G}%
_{2}$, in order to take their decision, need also to interact with their
environments $\mathcal{R}_{1}$ and $\mathcal{R}_{2}$. Contrary to what
happens for the players, whose situation can be described in a simple
four-dimensional Hilbert space, these environments are naturally defined in
an infinite-dimensional Hilbert space. For this reason, they can be thought
to describe some sub-system with infinite (or very many) degrees of freedom,
as the neurons in the brain, for instance. If we adopt this interpretation, $%
\mathcal{R}_{j}$ can be seen as the neural system of $\mathcal{G}_{j}$, $%
j=1,2$.

The full Hamiltonian $H$, see \cite{Bagarello2015a}, is the following:

\begin{equation}
\left\{
\begin{array}{ll}
H=H_0+H_I+H_{int}, &  \\
H_{0}=\sum_{j=1}^{2}\omega _{j}b_j^\dagger b_j+\sum_{j=1}^{2}\int_{\mathbb{R}%
}\Omega_j(k)B_j^\dagger(k)B_j(k)\,dk, &  \\
H_{I}=\sum_{j=1}^{2}\lambda_j\int_{\mathbb{R}}\left(b_j
B_j^\dagger(k)+B_j(k)b_j^\dagger\right)\,dk &  \\
H_{int}=\mu_{ex}\left(b_1^\dagger b_2+b_2^\dagger
b_1\right)+\mu_{coop}\left(b_1^\dagger b_2^\dagger+b_2 b_1\right). &
\end{array}%
\right.  \label{24}
\end{equation}

Here $\omega _{j}$, $\lambda _{j}$, $\mu _{ex}$ and $\mu _{coop}$ are real
quantities, and $\Omega _{j}(k)$ are real functions. In analogy with the $%
b_{j}$'s, we use fermionic operators $B_{j}(k)$ and $B_{j}^{\dagger }(k)$ to
describe the environment:
\begin{equation}
\{B_{i}(k),B_{l}(q)^{\dagger }\}=\delta _{i,l}\delta (k-q)\,1\!\!1,\qquad
\{B_{i}(k),B_{j}(k)\}=0,  \label{23}
\end{equation}%
which have to be added to those in (\ref{21}). Moreover each $b_{j}^{\sharp }
$ anti-commutes with each $B_{j}^{\sharp }(k)$: $\{b_{j}^{\sharp
},B_{l}^{\sharp }(k)\}=0$ for all $j$, $l$ and $k$. Here $X^{\sharp }$
stands for $X$ or $X^{\dagger }$. The various terms of $H$ can be understood
as follows: (1) $H_{0}$ is the \emph{free} Hamiltonian, which produces no
time evolution for the strategy operators $\hat{n}_{j}$. This is because $%
[H_{0},\hat{n}_{j}]=0$, and because $H$ reduces to $H_{0}$ in the absence of
interactions (i.e. when $\lambda _{j}=\mu _{ex}=\mu _{coop}=0$). This is in
agreement with our idea that the strategies of $\mathcal{G}_{1}$ and $%
\mathcal{G}_{2}$ can be modified only in the presence of interactions. (2) $%
H_{I}$ describes the interaction between the players and their neural
systems. Of course, the one discussed here is a special kind of interaction,
which is useful since it produces an analytical solution for the time
evolution of (the mean values of) the strategy operators. Other choices
could be considered, but these would, quite likely, break down this nice
aspect of the model. (3) $H_{int}$ describes two different interactions
between $\mathcal{G}_{1}$ and $\mathcal{G}_{2}$ . When $\mu _{coop}=0$ the
two players act differently, while they behave in the same way when $\mu
_{ex}=0$. Of course, when both $\mu _{coop}$ and $\mu _{ex}$ are not zero,
the dynamics are even richer. We refer to \cite{Bagarello2015a} for more
details on (\ref{24}).

The Heisenberg equations of motion $\dot{X}(t)=i[H,X(t)]$ can now be deduced
by using the CAR (\ref{21}) and (\ref{23}) and using $H$ given in (\ref{24}%
):
\begin{equation}
\left\{
\begin{array}{ll}
\dot{b}_{1}(t)=-i\omega _{1}b_{1}(t)+i\lambda _{1}\int_{\mathbb{R}%
}B_{1}(k,t)\,dk-i\mu _{ex}b_{2}(t)-i\mu _{coop}b_{2}^{\dagger }(t), &  \\
\dot{b}_{2}(t)=-i\omega _{2}b_{2}(t)+i\lambda _{2}\int_{\mathbb{R}%
}B_{2}(k,t)\,dk-i\mu _{ex}b_{1}(t)+i\mu _{coop}b_{1}^{\dagger }(t), &  \\
\dot{B}_{j}(k,t)=-i\Omega _{j}(k)B_{j}(k,t)+i\lambda _{j}b_{j}(t), &
\end{array}%
\right.  \label{25}
\end{equation}%
$j=1,2$. The solution of this system of equations has been found similarly
to \cite{Bagarello2015a}, and it looks like:
\begin{equation}
b(t)=e^{i\,U\,t}b(0)+i\int_{0}^{t}e^{i\,U\,(t-t_{1})}\,\beta (t_{1})\,dt_{1},
\label{28}
\end{equation}%
where we have introduced the following quantities:
\begin{equation*}
b(t)=\left(
\begin{array}{c}
b_{1}(t) \\
b_{2}(t) \\
b_{1}^{\dagger }(t) \\
b_{2}^{\dagger }(t) \\
\end{array}%
\right) ,\,\beta (t)=\left(
\begin{array}{c}
\lambda _{1}\beta _{1}(t) \\
\lambda _{2}\beta _{2}(t) \\
-\lambda _{1}\beta _{1}^{\dagger }(t) \\
-\lambda _{2}\beta _{2}^{\dagger }(t) \\
\end{array}%
\right) ,\,U=\left(
\begin{array}{cccc}
i\nu _{1} & -\mu _{ex} & 0 & -\mu _{coop} \\
-\mu _{ex} & i\nu _{2} & \mu _{coop} & 0 \\
0 & \mu _{coop} & i\overline{\nu _{1}} & \mu _{ex} \\
-\mu _{coop} & 0 & \mu _{ex} & i\overline{\nu _{2}}%
\end{array}%
\right) ,
\end{equation*}%
and where $\Omega _{j}(k)=\Omega _{j}k$, $\Omega _{j}>0$, $\nu _{j}=i\omega
_{j}+\pi \frac{\lambda _{j}^{2}}{\Omega _{j}}$ and $\beta _{j}(t)=\int_{%
\mathbb{R}}B_{j}(k)e^{-i\Omega _{j}kt}\,dk$, $j=1,2$.

As already stated, the next step consists in taking the average of the time
evolution of the strategy operators, $\hat{n}_{j}(t)=b_{j}^{\dagger
}(t)b_{j}(t)$, on a state over the full system $\mathcal{S}=\mathcal{S}_{%
\mathcal{G}}\otimes \mathcal{R}$, where $\mathcal{S}_{\mathcal{G}}$ has
already been introduced, $\mathcal{S}_{\mathcal{G}}=\{\mathcal{G}_{1},%
\mathcal{G}_{2}\}$, and $\mathcal{R}=\{\mathcal{R}_{1},\mathcal{R}_{2}\}$.
These states are assumed to be tensor products of vector states for $%
\mathcal{S}_{\mathcal{G}}$ and states on the environment in the following
way: for each operator of the form $X_{\mathcal{S}}\otimes Y_{\mathcal{R}}$,
$X_{\mathcal{S}}$ being an operator of $\mathcal{S}_{\mathcal{G}}$ and $Y_{%
\mathcal{R}}$ an operator of the environment, we have
\begin{equation*}
\left\langle X_{\mathcal{S}}\otimes Y_{\mathcal{R}}\right\rangle
:=\left\langle \Psi _{0},X_{\mathcal{S}}\Psi _{0}\right\rangle \,\omega _{%
\mathcal{R}}(Y_{\mathcal{R}}).
\end{equation*}%
Here $\Psi _{0}$ is the vector introduced in (\ref{20}), while $\omega _{%
\mathcal{R}}(.)$ is a state satisfying the following standard properties,
see \cite{Barnett1997}:
\begin{equation}
\omega _{\mathcal{R}}(1\!\!1_{\mathcal{R}})=1,\quad \omega _{\mathcal{R}%
}(B_{j}(k))=\omega _{\mathcal{R}}(B_{j}^{\dagger }(k))=0,\quad \omega _{%
\mathcal{R}}(B_{j}^{\dagger }(k)B_{l}(q))=N_{j}\,\delta _{j,l}\delta (k-q),
\label{29}
\end{equation}%
for some constant $N_{j}$. Also, $\omega _{\mathcal{R}}(B_{j}(k)B_{l}(q))=0$%
, for all $j$ and $l$.

\vspace{2mm}

\textbf{Remark:} At first sight, this expression for the state on $\mathcal{S%
}$ introduces a sort of asymmetry between $\mathcal{S}_{\mathcal{G}}$ and $%
\mathcal{R}$, since their states look of a different nature: the one over $%
\mathcal{S}_{\mathcal{G}}$ is a vector state, whilst the one over $\mathcal{R%
}$ is a linear positive functional over the algebra of the fermionic
operators $B_{j}(k)$ and $B_{j}^{\dagger }(k)$. This is not really
surprising, since $\mathcal{S}_{\mathcal{G}}$ involves just two degrees of
freedom, while $\mathcal{R}$ involves an infinite number of degrees of
freedom. Nevertheless, using the so-called
Gelfand-Naimark-Segal-construction (see \cite{Bratteli1987}), it is possible
to present $\omega _{\mathcal{R}}$ as a vector state.

\vspace{2mm}

After a few computations, calling $V(t)=e^{i\,U\,t}$ and $V_{k,l}(t)$ its $%
(k,l)$-matrix element, we deduce the following general formulas for the
\emph{Decision functions} (DFs) of $\mathcal{G}_{1}$ and $\mathcal{G}_{2}$,
which extend those found in \cite{Bagarello2015a} in absence of interference:

\begin{equation}
\left\{
\begin{array}{ll}
n_{1}(t)=\left\langle \left( b_{1}^{\dagger }(t)b_{1}(t)\right) ^{\dagger
}\right\rangle =\mu _{1}^{(\mathcal{G})}(t)+\delta \mu _{1}^{(\mathcal{G}%
)}(t)+n_{1}^{(B)}(t), &  \\
n_{2}(t)=\left\langle \left( b_{2}^{\dagger }(t)b_{2}(t)\right) ^{\dagger
}\right\rangle =\mu _{2}^{(\mathcal{G})}(t)+\delta \mu _{2}^{(\mathcal{G}%
)}(t)+n_{2}^{(B)}(t). &
\end{array}%
\right.   \label{230}
\end{equation}%
Here, we have introduced \ \ \
\begin{equation}
\left\{
\begin{array}{ll}
\mu _{1}^{(\mathcal{G})}(t)=\left\vert V_{1,1}(t)\right\vert ^{2}\left(
\left\vert \alpha _{1,0}\right\vert ^{2}+\left\vert \alpha _{1,1}\right\vert
^{2}\right) +\left\vert V_{1,2}(t)\right\vert ^{2}\left( \left\vert \alpha
_{0,1}\right\vert ^{2}+\left\vert \alpha _{1,1}\right\vert ^{2}\right) + &
\\
\qquad \qquad +\left\vert V_{1,3}(t)\right\vert ^{2}\left( \left\vert \alpha
_{0,0}\right\vert ^{2}+\left\vert \alpha _{0,1}\right\vert ^{2}\right)
+\left\vert V_{1,4}(t)\right\vert ^{2}\left( \left\vert \alpha
_{0,0}\right\vert ^{2}+\left\vert \alpha _{1,0}\right\vert ^{2}\right)  &
\\
\mu _{2}^{(\mathcal{G})}(t)=\left\vert V_{2,1}(t)\right\vert ^{2}\left(
\left\vert \alpha _{1,0}\right\vert ^{2}+\left\vert \alpha _{1,1}\right\vert
^{2}\right) +\left\vert V_{2,2}(t)\right\vert ^{2}\left( \left\vert \alpha
_{0,1}\right\vert ^{2}+\left\vert \alpha _{1,1}\right\vert ^{2}\right) + &
\\
\qquad \qquad +\left\vert V_{2,3}(t)\right\vert ^{2}\left( \left\vert \alpha
_{0,0}\right\vert ^{2}+\left\vert \alpha _{0,1}\right\vert ^{2}\right)
+\left\vert V_{2,4}(t)\right\vert ^{2}\left( \left\vert \alpha
_{0,0}\right\vert ^{2}+\left\vert \alpha _{1,0}\right\vert ^{2}\right) , &
\end{array}%
\right.   \label{231}
\end{equation}

\begin{equation}
\left\{
\begin{array}{ll}
\delta \mu _{1}^{(\mathcal{G})}(t)=2\Re \left[ \overline{V_{1,1}(t)}%
\,V_{1,2}(t)\overline{\alpha _{1,0}}\,\alpha _{0,1}+\overline{V_{1,1}(t)}%
\,V_{1,4}(t)\overline{\alpha _{1,1}}\,\alpha _{0,0}\right] + &  \\
\qquad \qquad -2\Re \left[ \overline{V_{1,2}(t)}\,V_{1,3}(t)\overline{\alpha
_{1,1}}\,\alpha _{0,0}+\overline{V_{1,3}(t)}\,V_{1,4}(t)\overline{\alpha
_{0,1}}\,\alpha _{1,0}\right] , &  \\
\delta \mu _{2}^{(\mathcal{G})}(t)=2\Re \left[ \overline{V_{2,1}(t)}%
\,V_{2,2}(t)\overline{\alpha _{1,0}}\,\alpha _{0,1}+\overline{V_{2,1}(t)}%
\,V_{2,4}(t)\overline{\alpha _{1,1}}\,\alpha _{0,0}\right] + &  \\
\qquad \qquad -2\Re \left[ \overline{V_{2,2}(t)}\,V_{2,3}(t)\overline{\alpha
_{1,1}}\,\alpha _{0,0}+\overline{V_{2,3}(t)}\,V_{2,4}(t)\overline{\alpha
_{0,1}}\,\alpha _{1,0}\right] , &
\end{array}%
\right.   \label{232}
\end{equation}%
and
\begin{equation}
\left\{
\begin{array}{ll}
n_{1}^{(B)}(t)=2\pi \int_{0}^{t}dt_{1}\left[ \frac{\lambda _{1}^{2}}{\Omega
_{1}}\left(
|V_{1,1}(t-t_{1})|^{2}N_{1}+|V_{1,3}(t-t_{1})|^{2}(1-N_{1})\right) \right] +
&  \\
\qquad \qquad +2\pi \int_{0}^{t}dt_{1}\left[ \frac{\lambda _{2}^{2}}{\Omega
_{2}}\left(
|V_{1,2}(t-t_{1})|^{2}N_{2}+|V_{1,4}(t-t_{1})|^{2}(1-N_{4})\right) \right] ,
&  \\
n_{2}^{(B)}(t)=2\pi \int_{0}^{t}dt_{1}\left[ \frac{\lambda _{1}^{2}}{\Omega
_{1}}\left(
|V_{2,1}(t-t_{1})|^{2}N_{1}+|V_{2,3}(t-t_{1})|^{2}(1-N_{1})\right) \right] +
&  \\
\qquad \qquad +2\pi \int_{0}^{t}dt_{1}\left[ \frac{\lambda _{2}^{2}}{\Omega
_{2}}\left(
|V_{2,2}(t-t_{1})|^{2}N_{2}+|V_{2,4}(t-t_{1})|^{2}(1-N_{4})\right) \right] .
&
\end{array}%
\right.   \label{233}
\end{equation}%
In formula (\ref{230}) we have clearly divided contributions of three
different natures: $\mu _{j}^{(\mathcal{G})}(t)$ contains contributions only
due to the players $\mathcal{G}_{1}$ and $\mathcal{G}_{2}$. Their analytic
expressions become particularly simple if the initial state $\Psi _{0}$ is
just one of the vectors $\varphi _{j,k}$, i.e. if all the coefficients $%
\alpha _{k,l}$ in (\ref{20}) are zero, except one. In this particular
situation, all the contributions in $\delta \mu _{j}^{(\mathcal{G})}(t)$,
which again only ref \ er to $\mathcal{G}_{1}$ and $\mathcal{G}_{2}$, are
zero. For this reason, we call $\delta \mu _{1}^{(\mathcal{G})}(t)$ and $%
\delta \mu _{2}^{(\mathcal{G})}(t)$ \emph{interference terms}: they are only
present if $\Psi _{0}$ is some superposition of eigenvectors of the (time
zero) number operators. Otherwise, they simply disappear. Finally, $%
n_{1}^{(B)}(t)$ and $n_{2}^{(B)}(t)$ arise because of the interaction of the
players with the environments: as we see from (\ref{233}), they are both
zero if $\lambda _{1}$ and $\lambda _{2}$ in the Hamiltonian are both equal
to zero, and they do not depend on the explicit form of $\Psi _{0}$.

A detailed analysis of what happens when there are no interference terms can
be found in \cite{Bagarello2015a}, where the focus was mainly on the
asymptotic behavior of the DFs in absence of interferences. Here, on the
other hand, we want to analyze what happens when the interference does exist
already at $t=0$ (i.e. when more than just one coefficient $\alpha _{j,k}$
is non zero), and we are also interested in the behavior of the DFs for
finite time.

\section{Analysis of the results}

\label{sectnumres}

In the figures plotted in this section we will call $\mathcal{C}_{1}$ the
following choice of parameters of $H$: $\omega _{1}=1$, $\omega _{2}=2$, $%
\Omega _{1}=\Omega _{2}=0.1$, $\lambda _{1}=\lambda _{2}=0.5$, and $\mathcal{%
C}_{2}$ the second choice: $\omega _{1}=0.1$, $\omega _{2}=0.2$, $\Omega
_{1}=\Omega _{2}=1$, $\lambda _{1}=1$ and $\lambda _{2}=0.7$. Also, we call $%
\mathcal{C}_{\alpha ,1}$ and $\mathcal{C}_{\alpha ,2}$ the following choices
of the parameters $\alpha _{k,l}$ in (\ref{20}): $\mathcal{C}_{\alpha,1
}=\{\alpha _{k,l}=\frac{1}{2},\forall \,k,l\}$, while $\mathcal{C}_{\alpha,2
}=\{\alpha _{0,1}=\frac{1}{2}=-\alpha _{1,1},\,\alpha _{0,0}=\frac{i}{2}%
=-\alpha _{1,0}\}$.

\begin{figure}[th]
\begin{center}
\includegraphics[width=0.4\textwidth]{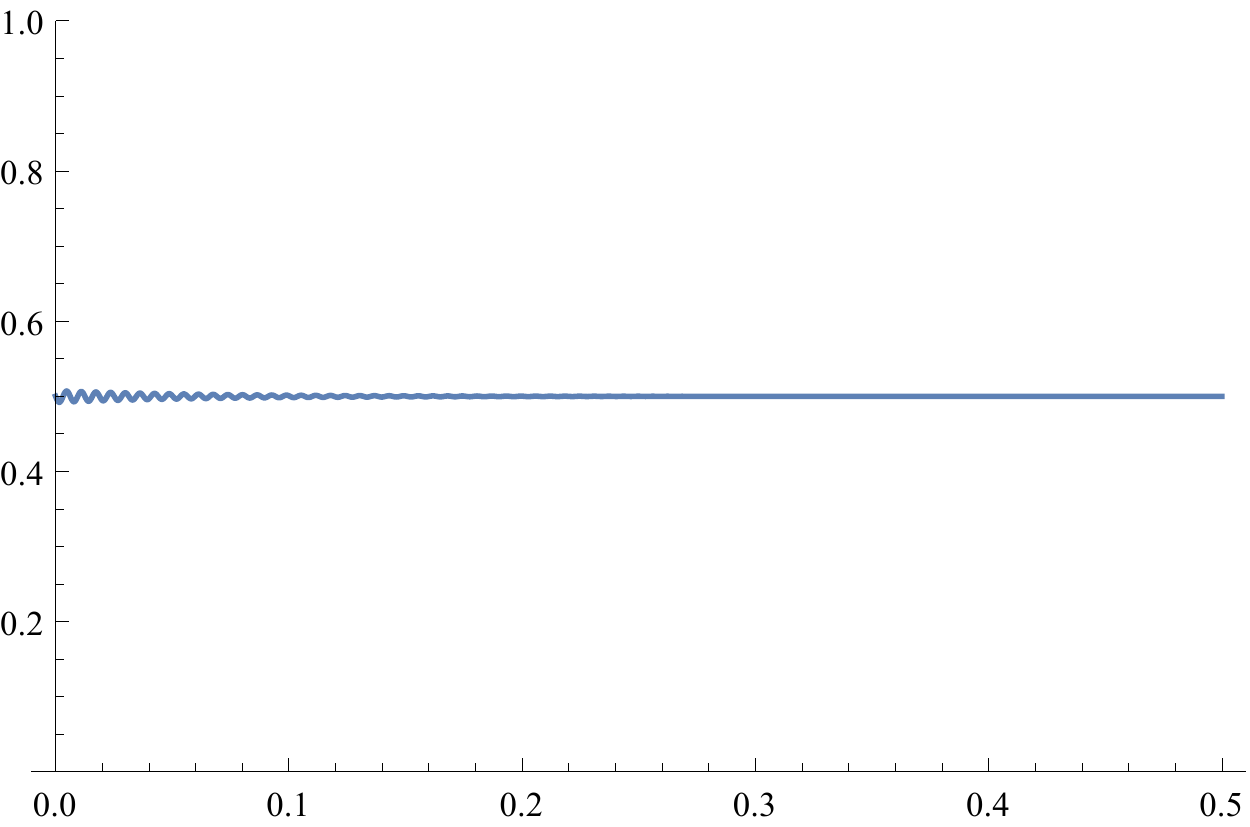}\hspace{%
8mm} \includegraphics[width=0.4\textwidth]{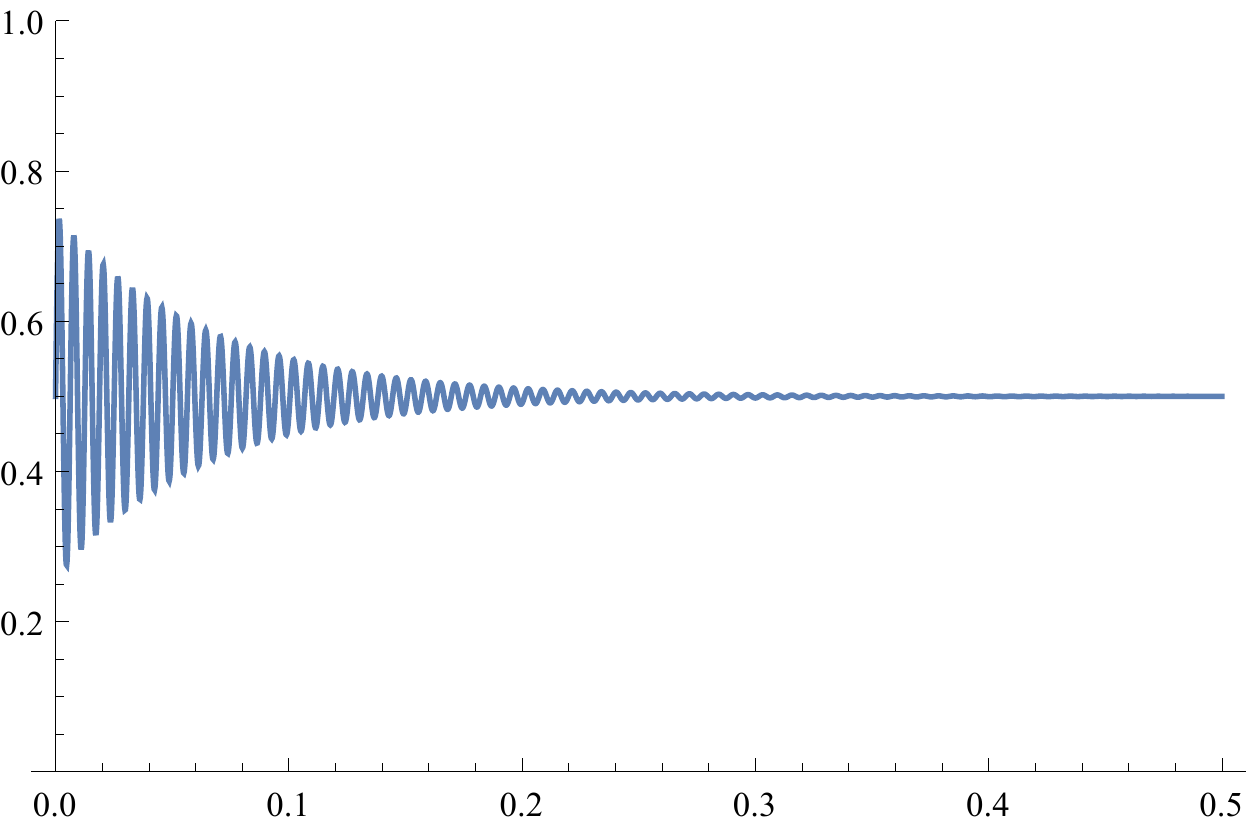}%
\hfill\\[0pt]
\includegraphics[width=0.4\textwidth]{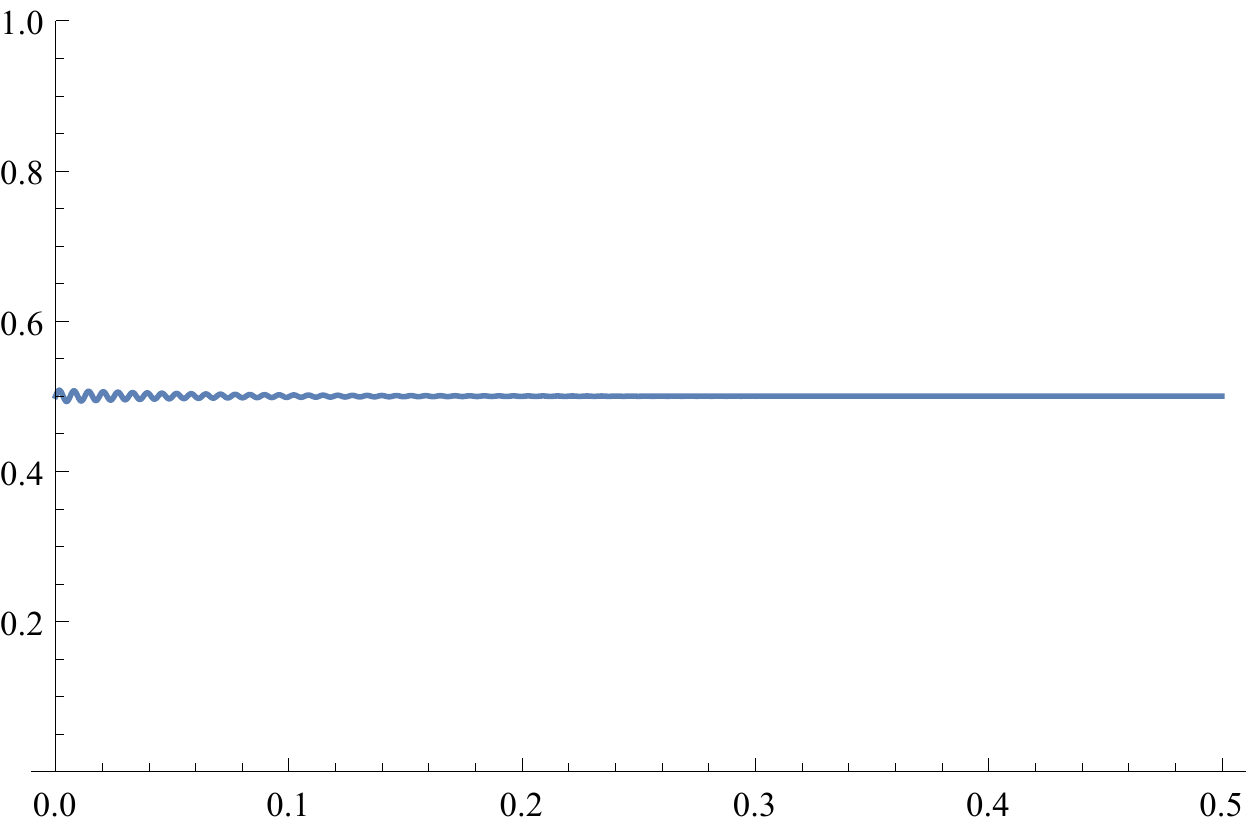}\hspace{%
8mm} \includegraphics[width=0.4\textwidth]{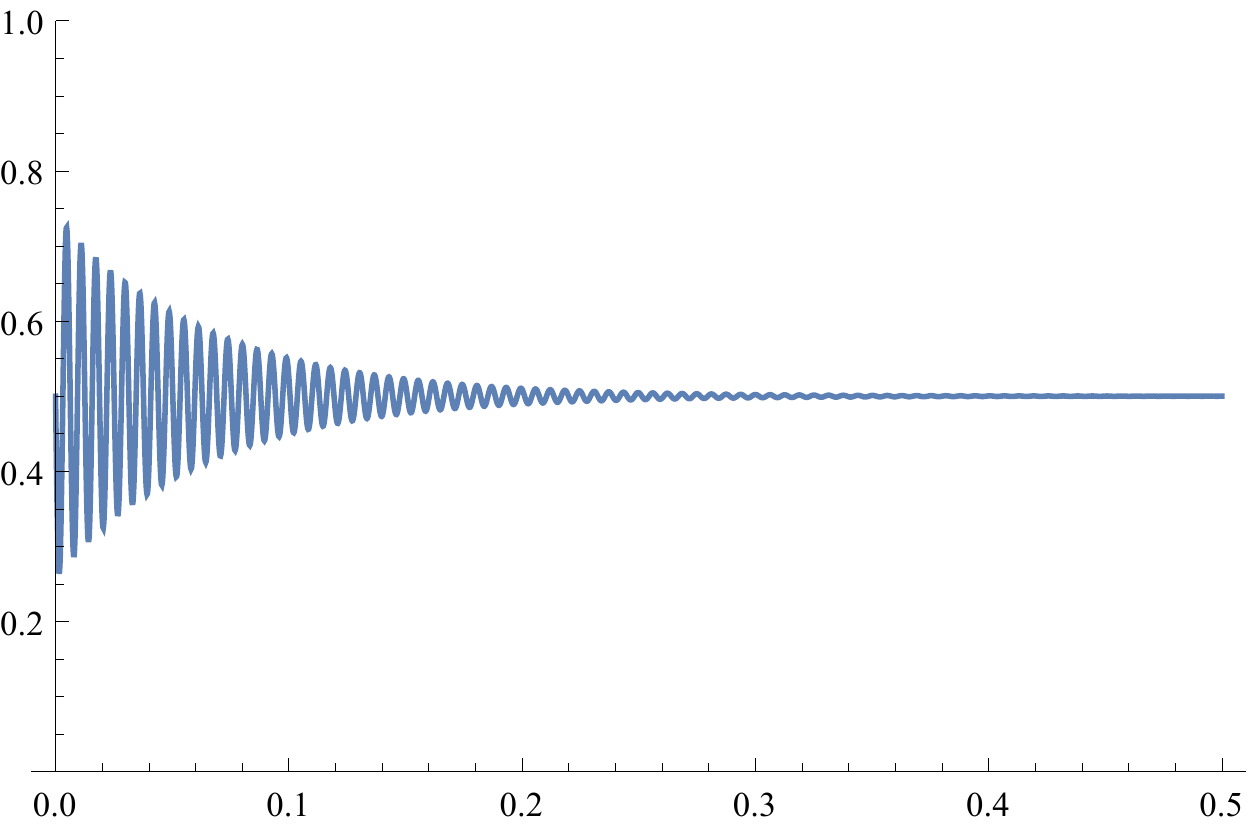}%
\hfill\\[0pt]
\end{center}
\caption{{\protect\footnotesize The DFs $n_{1}(t)$ (up) and $n_{2}(t)$
(down) for parameters $\mathcal{C}_{1}$, $N_{1}=0$, $N_{2}=1$, $\protect\mu %
_{ex}=500$, $\protect\mu _{coop}=0$ and for $\mathcal{C}_{\protect\alpha,1 }$
(left) and $\mathcal{C}_{\protect\alpha,2 }$ (right). }}
\label{fig1}
\end{figure}

Of course, several other choices could also be considered. However, the ones
we are fixing here cover already different situations. In particular, while
in $\mathcal{C}_{1}$ the interaction parameters $\lambda _{1}$ and $\lambda
_{2}$ are equal, they are different in $\mathcal{C}_{2}$. Also, while in $%
\mathcal{C}_{1}$ each $\omega _{j}$ is bigger than each $\Omega _{k}$, the
opposite holds for $\mathcal{C}_{2}$. This is important, since it is known
that the $\omega _{j} $'s are related to the inertia of the player $\mathcal{%
G}_{j}$ (see \cite{Bagarello2012}, \cite{Bagarello2015a}). Moreover, as for
the choices $\mathcal{C}_{\alpha ,1}$ and $\mathcal{C}_{\alpha ,2}$, the
difference is clear: in $\mathcal{C}_{\alpha ,1}$ all the coefficients in (%
\ref{20}) are equal, and in particular there is no relative phase between
the various $\varphi _{k,l}$'s in $\Psi_0$, whilst this is not so when
adopting the choice $\mathcal{C}_{\alpha ,2}$. And, as we will see, this
makes indeed a big difference: we clearly see that, even if (as expected
from what is deduced in \cite{Bagarello2015a}) the asymptotic values of the
two DFs appear to be independent of the choice of $\mathcal{C}_{\alpha,1 }$
and $\mathcal{C}_{\alpha,2 }$, there exists a certain time window in which
the choice of the $\alpha _{k,l}$'s really change the behaviors of the
functions. More concretely, if we add a phase in the coefficients defining
the original vector $\Psi _{0}$, we may observe quite large oscillations.
Then, \emph{interference terms in $\Psi _{0}$ make it, in general, quite
difficult to get a decision}. In particular, this is the effect of the
relative phases in the interference coefficients (see Figure \ref{fig1},
right), while if these coefficients have all the same phases, a decision can
be reached quite soon (see Figure \ref{fig1}, left). However, if we wait for
a sufficiently long time, in both cases we reach the same final values of
the DFs: the asymptotic values of the DFs only depend on the state of the
environment, and not on the particular choice of $\Psi _{0}$.

\begin{figure}[ht]
\begin{center}
\includegraphics[width=0.4\textwidth]{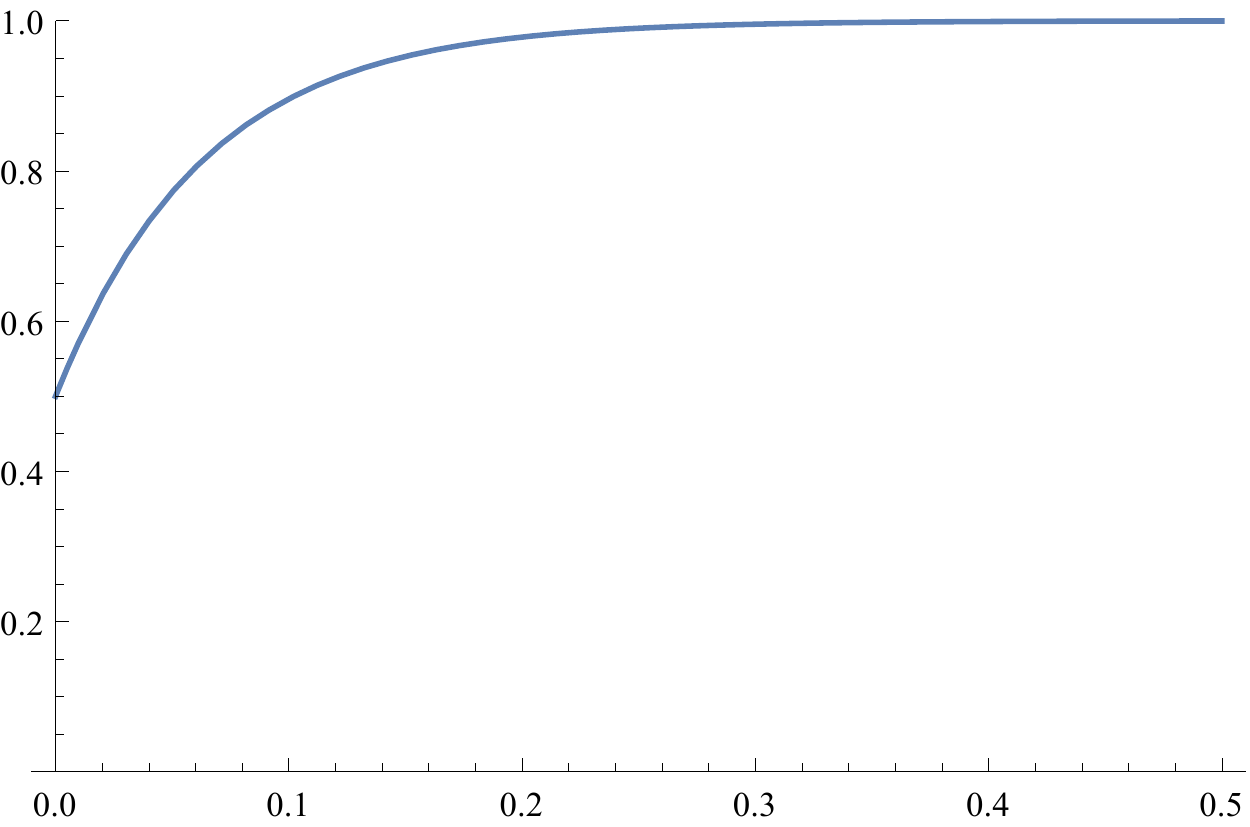}%
\hspace{8mm} \includegraphics[width=0.4%
\textwidth]{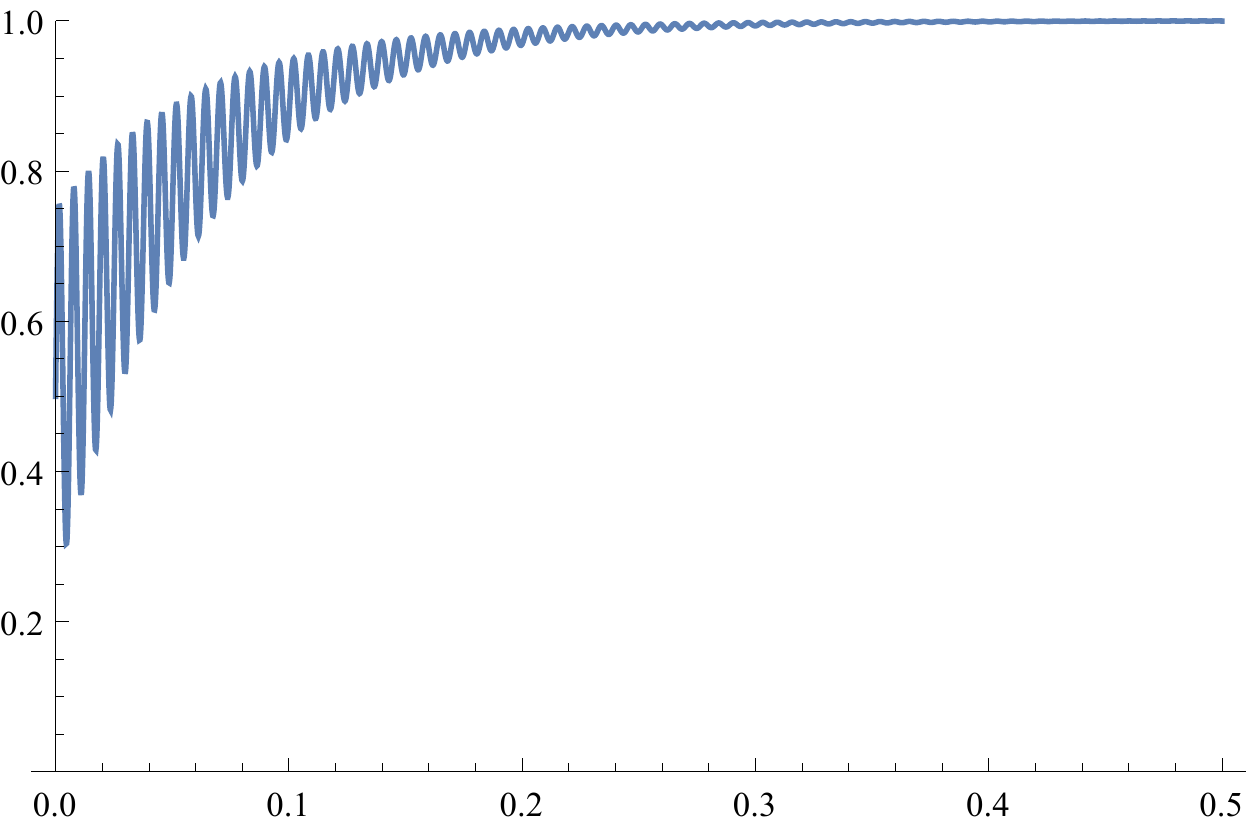}\hfill\\[0pt]
\includegraphics[width=0.4\textwidth]{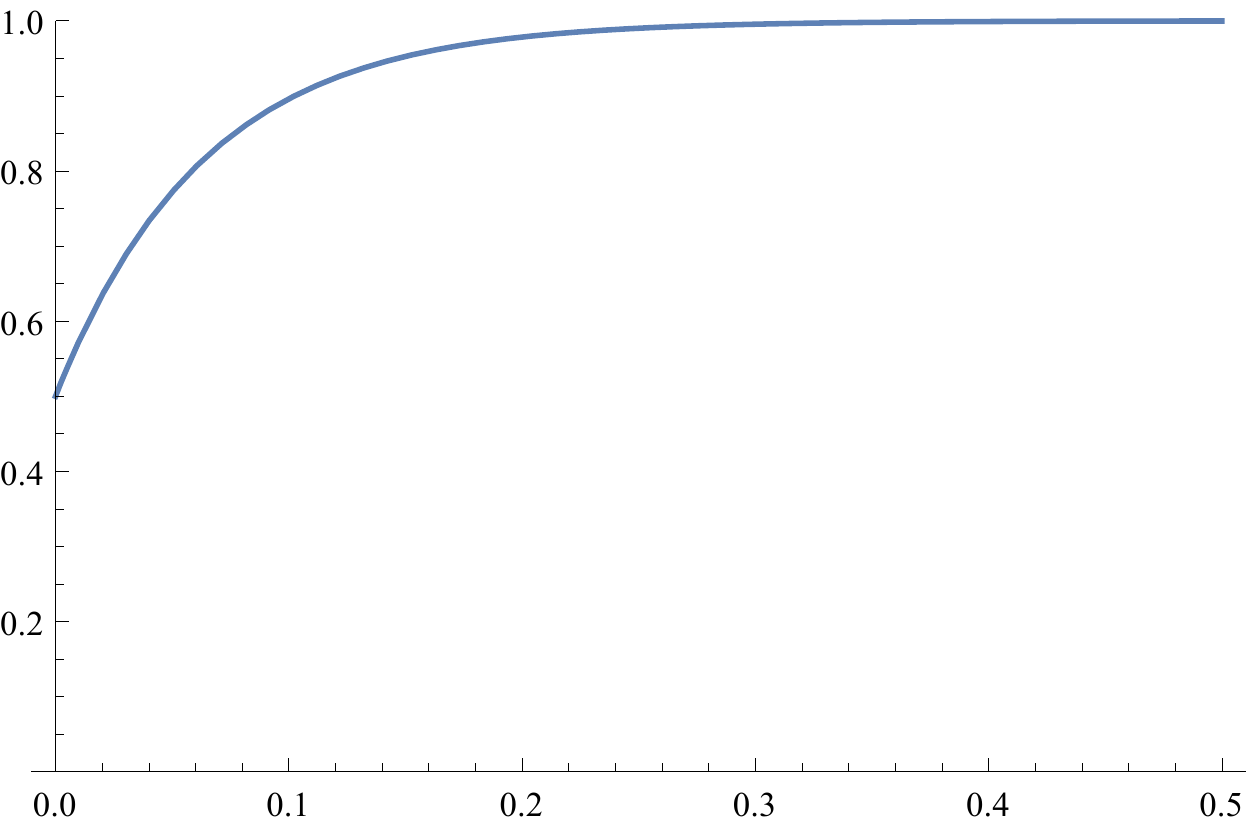}%
\hspace{8mm} \includegraphics[width=0.4%
\textwidth]{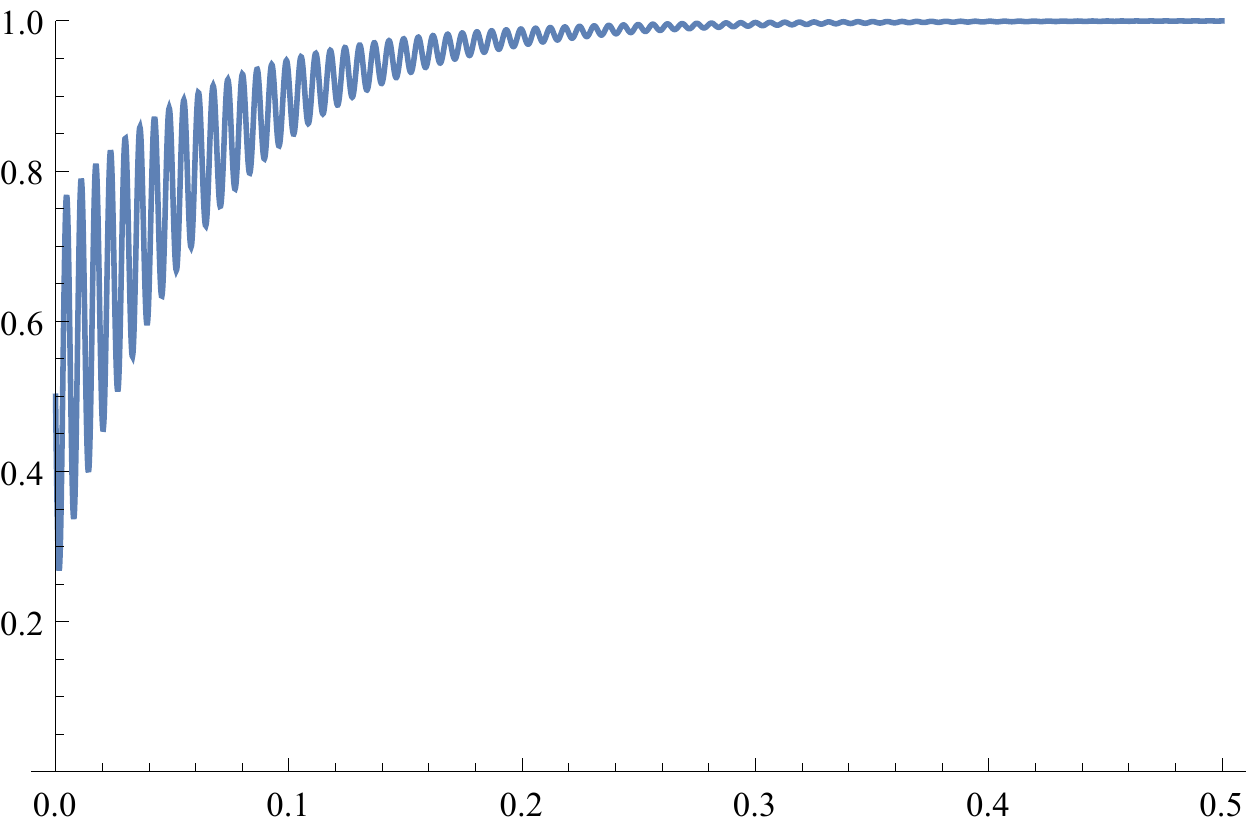}\hfill\\[0pt]
\end{center}
\caption{{\protect\footnotesize The DFs $n_1(t)$ (up) and $n_2(t)$ (down)
for parameters $\mathcal{C}_1$, $N_1=1$, $N_2=1$, $\protect\mu_{ex}=500$, $%
\protect\mu_{coop}=0$ and for $\mathcal{C}_{\protect\alpha,1}$ (left) and $%
\mathcal{C}_{\protect\alpha,2}$ (right). }}
\label{fig2}
\end{figure}

\begin{figure}[ht]
\begin{center}
\includegraphics[width=0.4\textwidth]{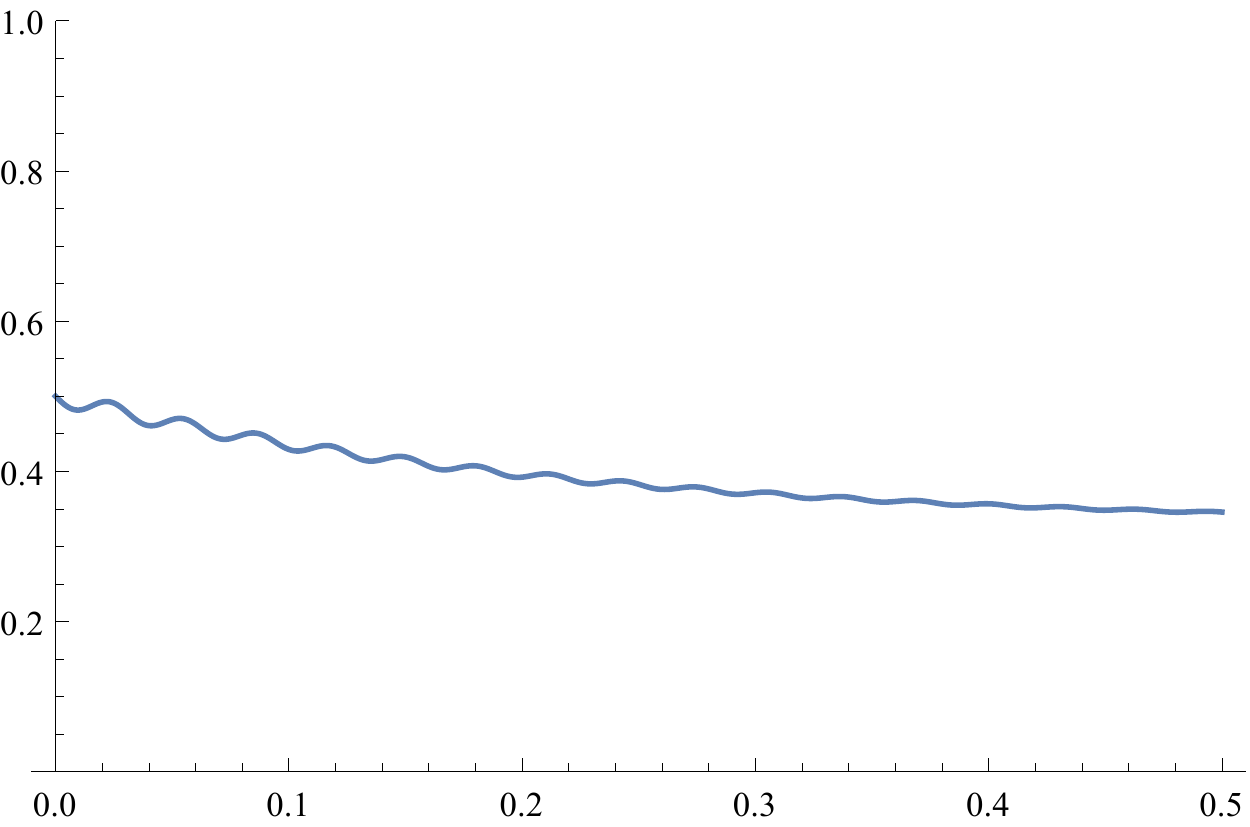}%
\hspace{8mm} \includegraphics[width=0.4%
\textwidth]{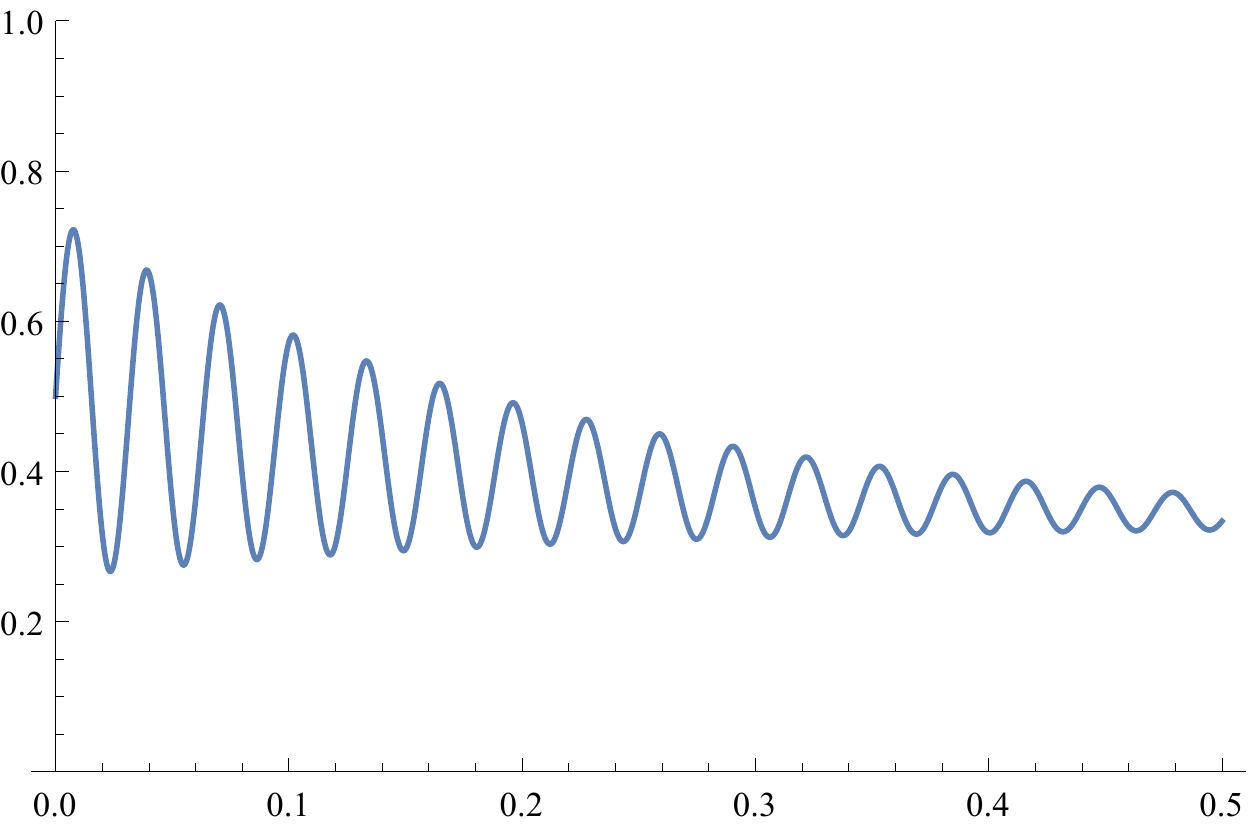}\hfill\\[0pt]
\includegraphics[width=0.4\textwidth]{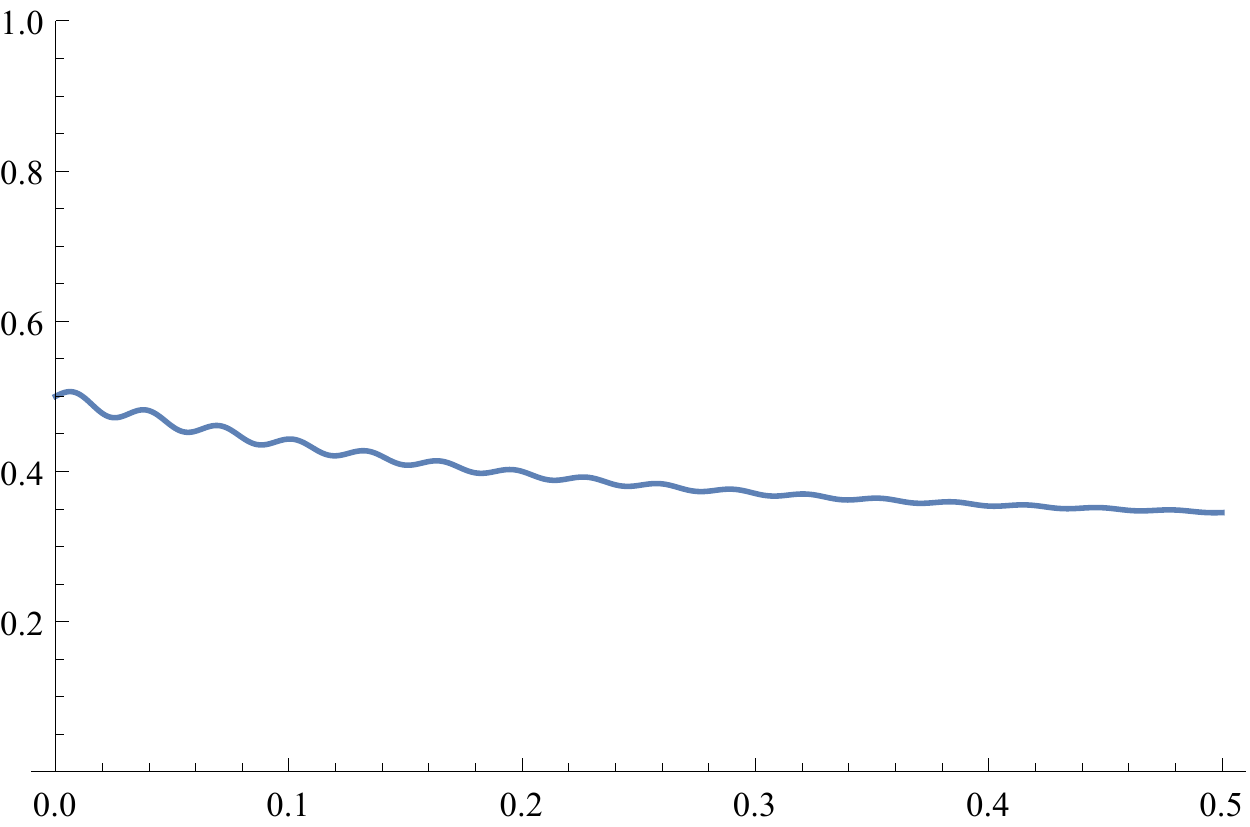}%
\hspace{8mm} \includegraphics[width=0.4%
\textwidth]{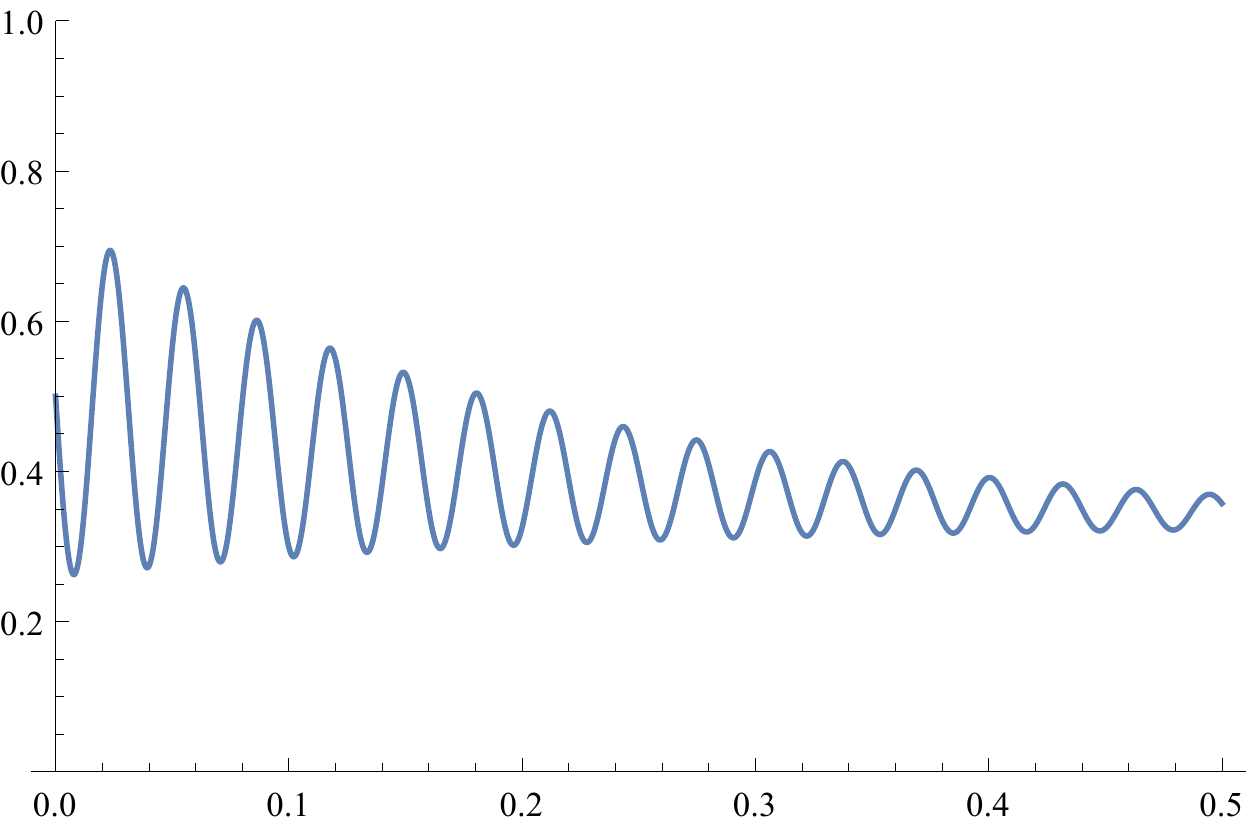}\hfill\\[0pt]
\end{center}
\caption{{\protect\footnotesize The DFs $n_1(t)$ (up) and $n_2(t)$ (down)
for parameters $\mathcal{C}_2$, $N_1=0$, $N_2=1$, $\protect\mu_{ex}=100$, $%
\protect\mu_{coop}=0$ and for $\mathcal{C}_{\protect\alpha,1}$ (left) and $%
\mathcal{C}_{\protect\alpha,2}$ (right). }}
\label{fig3}
\end{figure}

These conclusions are confirmed by other choices of the state on the
environment. For instance, in Figure \ref{fig2} we plot again the DFs for
the choices $\mathcal{C}_{\alpha,1 }$ (left) and $\mathcal{C}_{\alpha, 2 }$
(right) and for the same choice of the parameters as in Figure \ref{fig1},
while the state of the environment is chosen different, since we take now $%
N_{1}=N_{2}=1$. We see that there is no particular difference between the
two DFs $n_{1}(t)$ and $n_{2}(t)$ (this is possibly due to the fact that $%
N_{1}=N_{2}$). However, adding the phases to the $\alpha _{k,l}$ creates,
again, a lot of noise in the decision making process, noise which
disappears, but only after a sufficiently long time.

It is useful to stress that, changing further the parameters of the
Hamiltonian, does not really affect our conclusions. Indeed, Figure \ref%
{fig3} shows that even with different choices of the parameters in $H$, the
choice $\mathcal{C}_{2}$ with $\mu _{ex}=100$ (rather than $\mu _{ex}=500$,
as in the previous figures), phases create noise, and, again, this noise
becomes smaller and smaller after some time.

In this case, the existence of an asymptotic limit for the DFs is less
evident, but this is only due to the \emph{small} time interval considered.
It is not hard to imagine that we could still recover a clear asymptotic
value for $n_{j}(t)$ if we consider a time interval larger than just $%
[0,0.5] $. However, we will not do this here in order to keep range of the
figures uniform.

So far, we have taken $\mu _{coop}=0$. The same behavior is observed if we
take $\mu _{coop}\neq0$: we could check (but we do not include the plots
here) that the choice $\mathcal{C}_{\alpha,2}$ is again much more noisy than
$\mathcal{C}_{\alpha,1 }$.

\begin{figure}[ht]
\begin{center}
\includegraphics[width=0.4%
\textwidth]{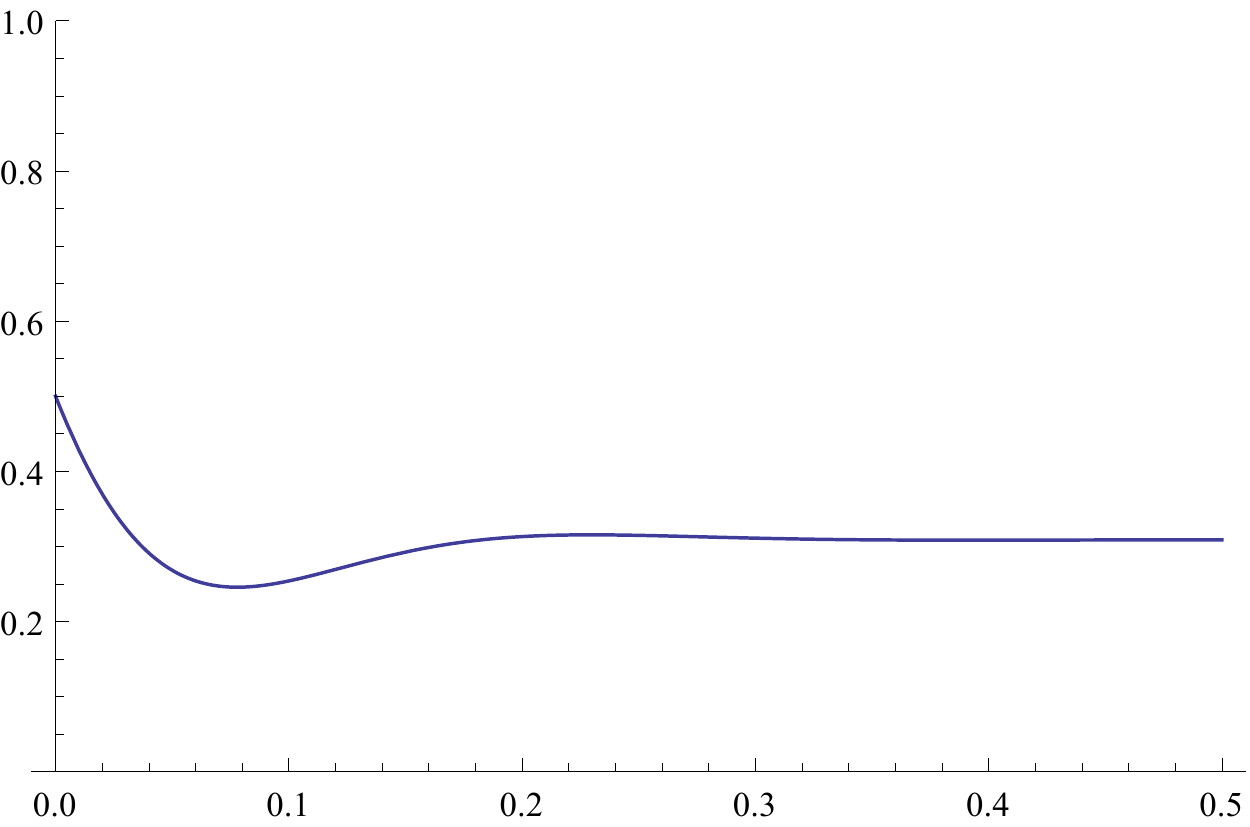}\hspace{8mm} %
\includegraphics[width=0.4%
\textwidth]{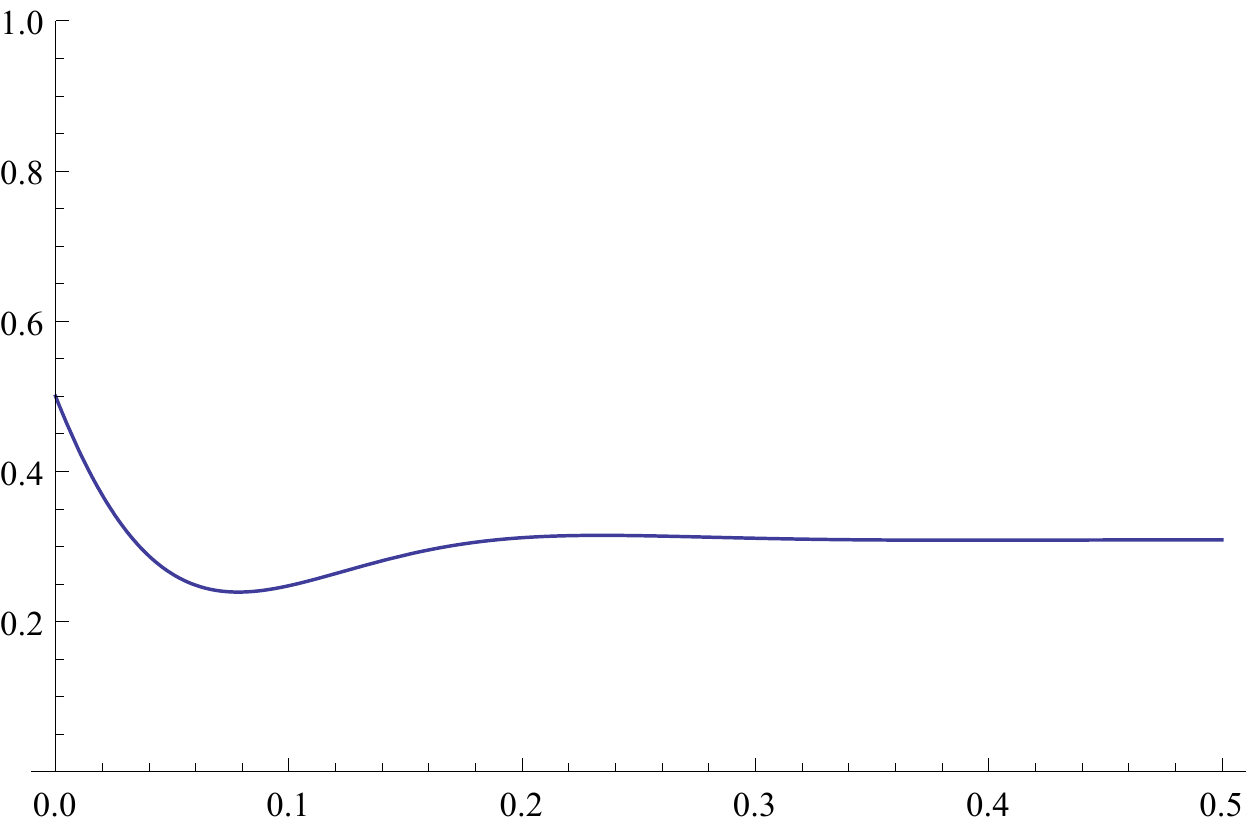}\hfill\\[0pt]
\includegraphics[width=0.4%
\textwidth]{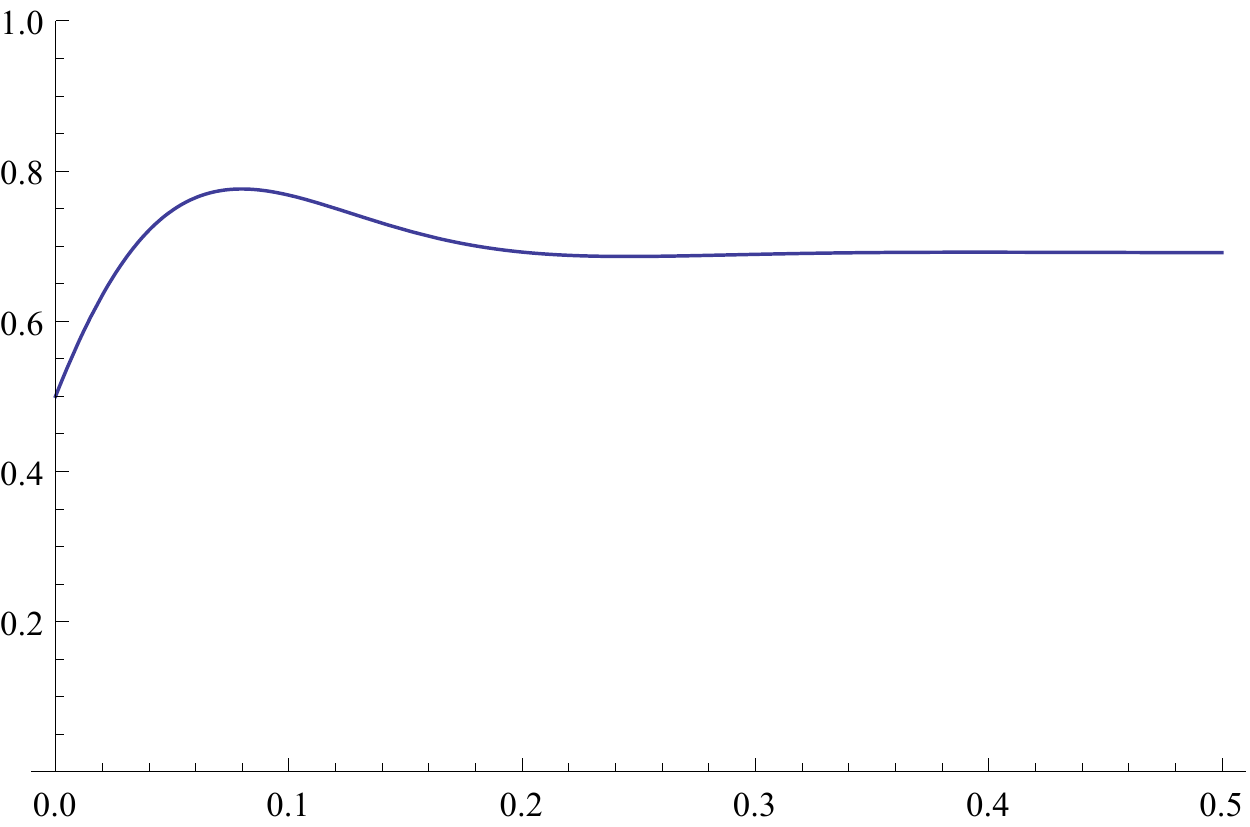}\hspace{8mm} %
\includegraphics[width=0.4%
\textwidth]{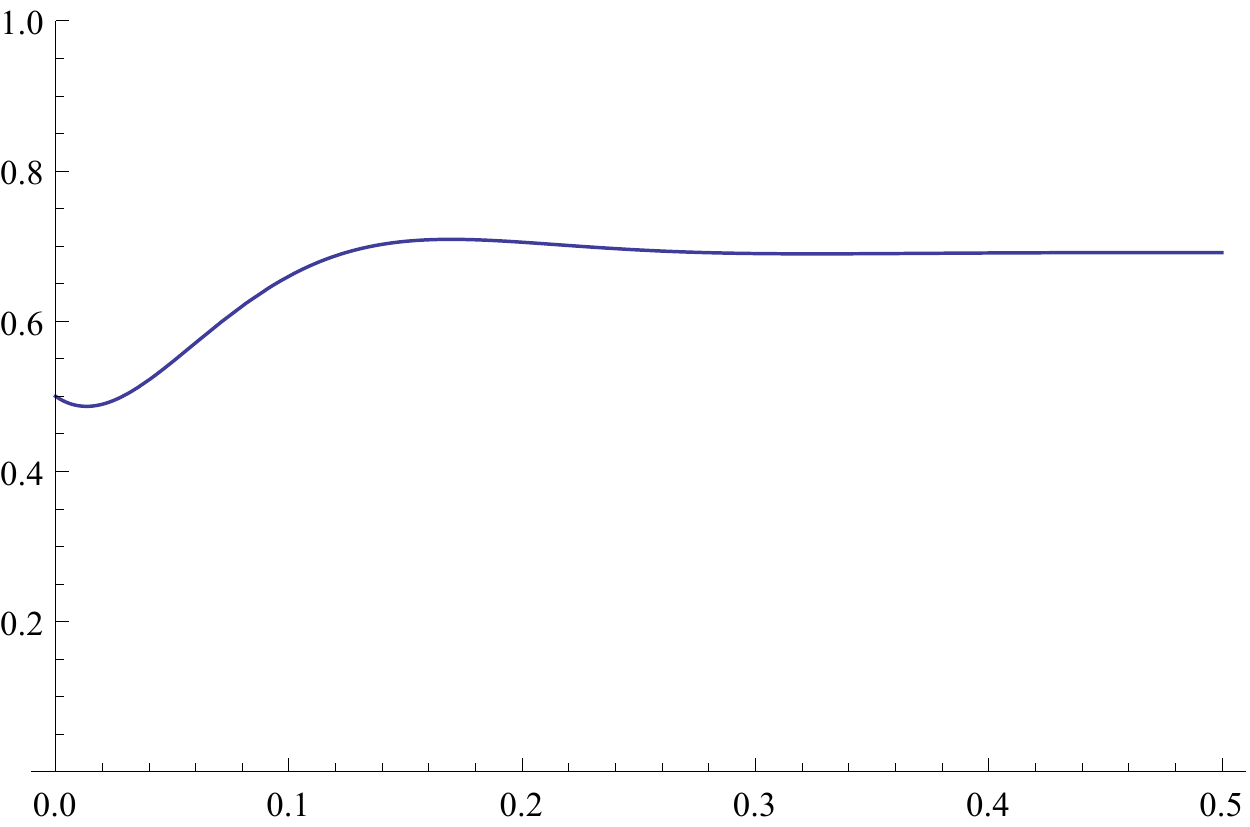}\hfill\\[0pt]
\end{center}
\caption{{\protect\footnotesize The DFs $n_1(t)$ (up) and $n_2(t)$ (down)
for parameters $\mathcal{C}_1$, $N_1=0$, $N_2=1$, $\protect\mu_{ex}=10$, $%
\protect\mu_{coop}=10$ and for $\mathcal{C}_{\protect\alpha,1}$ (left) and $%
\mathcal{C}_{\protect\alpha,2}$ (right). }}
\label{fig6}
\end{figure}

Also, no particular difference arises if we consider both $\mu_{ex}$ and $%
\mu_{coop}$ different from zero, especially when they are different enough
(again, we do not include the plots here). However, interestingly enough,
this noise is not so evident if $\mu_{ex}$ and $\mu_{coop}$ are not so
different, see Figure \ref{fig6}. It seems that, when the two terms in $%
H_{int}$, see (\ref{24}), act in cooperation, they become capable to \emph{%
filter the noise}, making the effect of the phases in $\alpha _{k,l}$ not so
strong. This is an interesting feature, which is surely worthy of a deeper
analysis.




\section{Discussion}

\label{INTS}

In this section we shall discuss the output of our model in detail.

\subsection{State's interpretation}

\label{SPR}

As is well known, the present situation in quantum foundations is
characterized by a huge diversity of interpretations of the quantum state.
The two main classes of interpretations correspond to \textit{the
statistical and individual viewpoints on a quantum state }(see Khrennikov
\cite{Khrennikov2016b}) for details. By the former a quantum state encodes
probabilities for the results of measurements for an ensemble of identically
prepared quantum systems.

We investigated the behavior of the averages of the strategy operators of
the two players $\mathcal{G}_{1}$ and $\mathcal{G}_{2}$ interacting with the
corresponding environments (`mental baths') $\mathcal{R}_{1}$ and $\mathcal{R%
}_{2}.$ One of the main outputs of our quantum-like model is that these
averages stabilize for $t\rightarrow \infty $ (for natural Hamiltonians
describing the interaction between players and their interactions with
mental baths). To interpret this result, we have to fix one of the
interpretations of a quantum state.

The straightforward interpretation of this stabilization result can be
presented on the basis of the statistical interpretation of quantum states.
Here we consider a very large ensemble of pairs of players and the averages
of the strategy operators are treated as \textit{the ensemble averages.} Our
stabilization result implies that on average (with respect to this ensemble)
the strategies of players stabilize to fixed values in the very long run of
the decision making reflections (this is encoded in creation and
annihilation operators forming the Hamiltonian of the game).

Now we switch to the individual interpretation of a quantum state. Here the
question of the interpretation of probabilities encoded in a quantum state
is more complicated than in the case of the statistical interpretation. It
seems that, for our applications of the mathematical formalism of quantum
mechanics to the operational modeling of decision making, the most natural
interpretation of probability is \textit{the subjective interpretation.}
Recently this interpretation became popular in quantum information theory
and it is known as QBism (Quantum Bayesianism) (see \cite{Fuchs2011}, \cite%
{Fuchs2013})\footnote{%
We do not claim that QBism is the proper interpretation of quantum \textit{%
physics. }See Khrennikov \cite{Khrennikov2016b} for reflections from one of
the authors of this paper on the applicability of QBism in quantum physics.
But it is very natural to use it for our purpose (see \cite{Haven2017}; \cite%
{Khrennikov2016b}) for a motivation. We also remark that originally
classical game theory and the theory of decision making was formulated by
von Neumann and Morgenstern \cite{vonNeumann1944} in the statistical
probabilistic framework: probabilities were treated from the frequentist
viewpoint. However, Savage \cite{Savage1954} reformulated it by using the
subjectivist viewpoint on probability. Nowadays the latter is dominating the
foundations of decision making and the axiomatic foundation of economics,
but with the strong emphasize of the role of Bayesian inference.}.

In the QBist framework the average stabilization output of our model can be
interpreted in the following way:

At the initial instant of time $t=0,$ each pair of players assigns their
individual subjective probabilities to selections of their strategies,
\begin{equation*}
p_{i}(j)\equiv p_{i}(j;0),i=1,2,j=0,1.
\end{equation*}%
These probabilities depend on the initial state $\Psi _{0}$ of the pair of
players. By using the representation of $\Psi _{0}$ as superposition of the
`certainty states' $\varphi _{k,l},$ see (\ref{20}), and Born's rule, these
probabilities can be represented as
\begin{equation*}
p_{1}(j)=|\alpha _{j,0}|^{2}+|\alpha _{j,1}|^{2},\qquad p_{2}(j)=|\alpha
_{0,j}|^{2}+|\alpha _{1,j}|^{2},
\end{equation*}%
for $j=0,1$. In fact, for instance, we have $p_{1}(0)=|\left\langle \varphi
_{00},\Psi _{0}\right\rangle |^{2}+|\left\langle \varphi _{01},\Psi
_{0}\right\rangle |^{2}$, and so on. If $\Psi _{0}$ is non-factorizable
(entangled), then, $p_{1}(j)$ cannot be expressed solely in terms of the
state of $\mathcal{G}_{1}$ and vice versa. We emphasize that in the case of
entanglement, for a pure state $\Psi _{0},$ the states of players are mixed
states and they are represented by density operators.

Starting with the initial state $\Psi _{0},$ and hence with the
probabilities $p_{i}(j),$ each decision maker updates continuously her state
by taking into account the impact of the environment and the feedback from
the update of the state of the co-player. This state dynamics studied in
previous sections generates the corresponding dynamics of the subjective
probabilities, $t\rightarrow p_{i}(j;t).$ The essence of exploring the
quantum-like model indicates that the genuine dynamics are not a dynamics of
probabilities, but rather a dynamics of the state. Metaphorically we can say
that the dynamics of probabilities are a shadow of the mental state
dynamics. The state dynamics can be expressed, see Section 3, in terms of
the strategy operators.

A direct computation shows that the strategy operators $\hat{n}_{i}$ (at $%
t=0 $) can be represented in the form (spectral decomposition):
\begin{equation}
\hat{n}_{1}=|\varphi _{1,0}\rangle \langle \varphi _{1,0}|+|\varphi
_{1,1}\rangle \langle \varphi _{1,1}|;  \label{Odds2}
\end{equation}%
\begin{equation}
\hat{n}_{2}=|\varphi _{0,1}\rangle \langle \varphi _{0,1}|+|\varphi
_{1,1}\rangle \langle \varphi _{1,1}|,  \label{Odds2a}
\end{equation}%
where $(|f\rangle\langle g|)h=\left\langle g,h\right\rangle \,f$, for all $%
f,g,h\in \mathcal{H}$. The ones in (\ref{Odds2})-(\ref{Odds2a}) are
orthogonal projectors onto the subspaces $L_{1}$ with the basis $(\varphi
_{1,0},\varphi _{1,1})$ and $L_{2}$ with the basis $(\varphi _{0,1},\varphi
_{11}).$ Those are the specifics of the fermionic representation of the
reflection and strategy operators.

Now, with simple manipulations, we observe that
\begin{equation*}
p_{1}(1;t)=|\left\langle \varphi _{10},\Psi (t)\right\rangle
|^{2}+|\left\langle \varphi _{11},\Psi (t)\right\rangle |^{2}=\left\langle
\Psi (t),\left( |\varphi _{1,0}\rangle \langle \varphi _{1,0}|+|\varphi
_{1,1}\rangle \langle \varphi _{1,1}|\right) \Psi (t)\right\rangle =
\end{equation*}%
\begin{equation*}
=\left\langle \Psi (t),\hat{n}_{1}\Psi (t)\right\rangle =\left\langle \Psi
_{0},\hat{n}_{1}(t)\Psi _{0}\right\rangle ,
\end{equation*}%
moving from the Schr\"{o}dinger to the Heisenberg representation. Therefore,
the averages of these operators, $i=1,2,$ are equal to the probabilities to
select the decision $s=1$
\begin{equation}
p_{i}(1;t)=\langle \hat{n}_{i}(t)\rangle,\qquad i=1,2.  \label{Odds1}
\end{equation}%
These are nothing else than DFs considered in section \ref{TMOD}, see (\ref%
{230}). This clarifies the relation between the subjective probabilities and
the decision functions.

\subsection{The process of decision making: the subjective probability
viewpoint}

\label{INTS1}

Each decision maker starts with assignment of the subjective probabilities $%
p_{i}(1)=\langle \hat{n}_{i}\rangle ,p_{i}(0)=1-\langle \hat{n}_{i}\rangle ,$
where $\hat{n}_{i}=\hat{n}_{i}(0).$ Then she begins her reflections on the
possible selections of the strategies (adapted to the environment and
interactions with the other decision maker) and her subjective probabilities
fluctuate, see Sections \ref{TMOD}, \ref{sectnumres}.

We have seen that the structure  of fluctuations depends crucially on the
initial state $\Psi _{0}.$ In the process of decision making, the magnitude
of these fluctuations decreases and subjective probabilities stabilize to
the fixed values. These are the probabilities which are used by the decision
maker to make her choice. After the determination of her subjective
probabilities, she proceeds as a classical decision maker. For the instant
of time $t,$ the odds in favor of the decision labeled by 1 are given by the
proportion:
\begin{equation}
O_{i}(1;t)=\frac{p_{i}(1;t)}{p_{i}(0;t)}.  \label{Odds}
\end{equation}%
We consider the instance of time $\tau $ when the decision is made. This $%
\tau $ corresponds to a diminishing of the fluctuations to the minimal
level. In the theoretical model $\tau =\infty .$ But in reality a decision
maker cannot wait for an infinite time to make the decision. The decision
maker uses some threshold for fluctuations $\epsilon >0$ and the decision
instant $\tau $ is determined by $\epsilon $.

If $O_{i}(1;\tau )>1,$ then $\mathcal{G}_{i}$ makes the decision labeled by
1. If $O_{i}(1;\tau )<1,$ then she makes the opposite decision (if $%
O_{i}(1;\tau )=1,$ she either makes the decision randomly with probability
1/2 or she repeats the process of decision making with a modified initial
state $\Psi _{0}).$ We remark that in the absence of the mental baths, the
subjective probabilities assigned by a decision maker to the possible
strategies would fluctuate for ever, as deduced, for instance, in \cite%
{Bagarello2015b} in a different context.

\subsection{Violation of the law of total probability: `interference of
probabilities'}

\label{INTS2}

As was emphasized by Feynman and Hibbs \cite{Feynman1965}, the probabilistic
data from interference experiments with quantum systems, e.g. the two slit
experiment, can be treated as violating the laws of classical probability.
They pointed to a violation of the additivity of probability. Feynman's
argument was reformulated by one of the coauthors of this paper in terms of
conditional probabilities - as a violation of the law of total probability
(LTP) \cite{Khrennikov2010}. In this form, this argument is nicely
applicable to the process of decision making. The classical Bayesian scheme
of decision making is fundamentally based on the LTP. Its violation leads to
non-Bayesian schemes of decision making and the quantum formalism provides
one of such schemes\footnote{%
Of course, a non-Bayesian scheme need not to be precisely quantum. It is not
clear whether the quantum formalism can cover all probabilistic phenomena
arising in decision making \cite{Boyer2015}, \cite{Boyer2016}.}.

We recall the LTP. Consider two random variable $\xi $ and $\eta $ taking
discrete values, $\xi =\alpha _{1},...,\alpha _{N}$ and $\eta =\beta
_{1},...,\beta _{M}.$ Then, in the classical probabilistic framework (the
measure-theoretic model of probability (see Kolmogorov, 1933) it can be
proven that:
\begin{equation}
p(\eta =\beta )=\sum_{\alpha }p(\xi =\alpha )p(\eta =\beta |\xi =\alpha ).
\label{PR0}
\end{equation}%
It is important to remark that the conditional probability is defined by the
Bayes formula: $p(\eta =\beta |\xi =\alpha )=p(\eta =\beta ,\xi =\alpha
)/p(\xi =\alpha ),$ where $p(\eta =\beta ,\xi =\alpha )$ is the joint
probability distribution of the pair of random variables. As is well known,
in general, for quantum observables, the joint probability is not well
defined. Therefore, the classical Bayes formula for conditional
probabilities is in-applicable and one can expect that, for quantum
conditional probabilities which are defined in the Hilbert space formalism,
the formula (\ref{PR0}) can be violated \cite{Khrennikov2010}.

The classical LTP in our setting can be formulated as follows. We remind
that the probabilities $p_{i}(s;t),s=0,1,$ depend on the initial state $\Psi
_{0},$ i.e., in fact, we could even use the notation $p_{i}(s;t|\Psi _{0}).$
Among the possible initial mental states of the pairs of players $(\mathcal{G%
}_{1},\mathcal{G}_{2}),$ the states $\varphi _{k,m},k,m=0,1,$ play the very
special role. Here initially the players were completely sure in their
strategies, $\mathcal{G}_{1}$ with the strategy $k$ and $\mathcal{G}_{2}$
with the strategy $m.$ In (\ref{PR0}) the probabilities $p_{i}(s;t|k,m)%
\equiv p_{i}(s;t|\varphi _{k,m}),s=0,1,$ can be identified with the
conditional probabilities $p(\eta =\beta |\xi =\alpha )$ with $\alpha =(k,m)$
and $\beta =0,1.$ The probabilities
\begin{equation}
p_{i}(k,m)=|\langle \Psi _{0}|\varphi _{k,m}\rangle |^{2},\;k,m=0,1,
\label{PR}
\end{equation}%
encode uncertainty in the determination of the pairs of strategies $(k,m)$
for the initial mental state $\Psi _{0}.$ In (\ref{PR0}) they are identified
with the probabilities $p(\xi =\alpha ).$

The interference-like representation of the decision functions (\ref{230})
can be rewritten as a quantum generalization of LTP:
\begin{equation}  \label{PR1}
p_i(s; t)= \sum_{(k,m)} p_i (k, m) p_i(s; t\vert k,m)
\end{equation}
\begin{equation*}
+ \sum_{(k,m) \not= (j,n) } \cos \theta_i(s; t\vert (k,m, j,n) \sqrt{p_i (k,
m) p_i(s; t\vert k,m) p_i (j, n) p_i(\alpha; t\vert j,n)}
\end{equation*}
Here the phase $\theta_i(s; t\vert (k,m, j,n)$ encodes a sort of
interference between the pairs of strategies $(k,m)$ and $(j, n).$ In our
quantum model this phase can be easily calculated by using the phases of the
states $\Psi_0$ and $\varphi_{k,m}, \varphi_{j,n}.$

As was shown in \cite{Khrennikov2010}, in the purely probabilistic framework
(i.e. without direct appealing to the Hilbert space formalism) a violation
of the LTP means the impossibility to embed all probabilities in (\ref{PR1}%
), into a single classical probability space. Thus, the appearance of the
additional interference-like term disturbing the classical LTP, see (\ref%
{PR1}), shows that classical probability has the restricted domain of
applications. For our applications to decision making, it is more useful to
interpret this result in terms of the interrelation between classical and
nonclassical logics. Classical probability theory is based on classical
Boolean logic. Therefore, the impossibility to use classical probability
implies the impossibility to use the Boolean logic. Thus, in our
quantum-like model the actions of the players are not ruled solely by the
laws of Boolean logics. They can make decisions with reasoning based on
non-classical logic.

\section{Conclusions and perspectives}

\label{sectconcl}

The DM under uncertainty plays a crucial role in economics, game theory and
many areas of social science. Until now, the most intensive and successful
modeling of the DM-process was performed on the basis of Bernoulli's idea to
explore utility functions (which originated in his attempting to solve the
St. Petersburg paradox). Several theories have ensued since then: expected
utility theory, prospect theory, cumulative prospect theory and other
approaches. In spite of the aforementioned success of these utility based
models, it is impossible to ignore the growing dissatisfaction by the
present state of the art in DM. In particular, the number of paradoxes have
increased over the years: Allais, Ellsberg, Machina and one recently counted
39 paradoxes (see \cite{Erev2016}). Already, in subjective uncertainty
utility models, the representation of the agent's utility by a function is a
fuzzy problem. Individuals may use a huge variety of possible utility
functions and the determination of the class of possible functions is a
complicated problem (see Machina \cite{Machina1982}, \cite{Machina1983},
\cite{Machina1987}).

One can treat DM as the interaction of an agent with a complex information
environment which includes a variety of behavioral, economic, social, and
geopolitical factors as well as beliefs about states of other agents (e.g.
the financial market). Therefore, it is natural to model DM as an
(environment-)adaptive dynamical process. In principle, one may try to
encode such a complex information environment by a single function encoding
the utility of a context. However, this would definitely be a high
simplification of the mathematical representation of the DM-context.

The most advanced mathematical model of interaction of a system with an
environment is presented in quantum theory. The modern quantum information
viewpoint on quantum theory (see Brukner and Zeilinger \cite{Brukner1999},
1999 and also Chiribella \cite{Chiribella2012} and Plotnitsky \cite%
{Plotnitsky2014}) justifies the possibility to apply this formalism outside
of physics. Here the keyword is \textit{adaptivity} to the environment. In
this paper we have applied the quantum model of adaptive dynamics to the
modeling of the DM-process. Our model is QFT-inspired: the information
environment is modeled with the aid of quantum fields which are treated as
operational quantities carrying information. In such a DM-model, an agent
does not maximize (expected) utility, but she searches for decisions
matching `demands'\ of the environment and (agent's representation) of the
belief-states of other agents involved in the DM-process. This dynamical
adaptive process is based on the representation of the DM-context by a
quantum pure state carrying maximal available information about the
situation.

We have found analytical and numerical solutions for operator-valued
functions representing the DM-process and their expectations, and we
considered the problem of stabilization for $t\rightarrow \infty .$ The
output of stabilization is considered as the classical (mixed) decision
strategy. Of course, the real DM-process cannot take an infinite time and we
are interested in approximate stabilization. We have carefully analyzed the
interpretational issues of our model (see section \ref{INTS}). In
particular, we have looked at the coupling between subjective probability
and the private agent interpretation of quantum mechanics (QBism) (see Fuchs
and Schack \cite{Fuchs2011}, \cite{Fuchs2013}).

This paper is of a conceptual nature. Its aim is to present a quite general
model of DM under uncertainty as an adaptive dynamical process of evolution
of the belief state of an agent interacting with a complex information
environment (e.g., the financial market). This paper also wants to
demonstrate the mathematical power of this general model. Our proposed model
can be explored for a variety of problems in economics, sociology, and
politics. See Bagarello \cite{Bagarello2015a}, \cite{Bagarello2015b}, \cite%
{Bagarello2016} and Khrennikova \cite{Khrennikova2016}\footnote{%
We remark that the QFT-inspired model generalizes the quantum dynamical
DM-model of Pothos and Busemeyer \cite{Pothos2009}. It also has close
connections with the dynamical DM-model based on the quantum master equation
(see Asano et al. \cite{Asano2011}, \cite{Asano2012a}, \cite{Asano2012b},
\cite{Asano2015}).} for concrete applications of special variations of the
presented model. Others are in progress.

\section*{Data accessibility statement}

This work does not have any experimental data.

\section*{Competing interests statement}

We have no competing interests.

\section*{Authors' contributions}

FB proposed the model and its solution. AK and EH worked on the
interpretation of the model in the context of DM.

\section*{Acknowledgements}

One of the authors (FB) acknowledges partial support from the University of
Palermo and from G.N.F.M. and the discussion on the paper was started during
his visit to Linnaeus University (supported by the  grant ``Mathematical
Modeling of Complex Hierarchic Systems''). FB also wishes to thank Prof. A.
Busacca for his strong support for this project.

\section*{Funding statement}

This work received partial financial support by the grant \textquotedblleft
Mathematical Modeling of Complex Hierarchic Systems\textquotedblright\ of
Linnaeus University and by the EU-project \textquotedblleft Quantum
Information Access and Retrieval Theory\textquotedblright\ (QUARTZ), Grant
No. 721321.


\end{document}